\documentclass{aa}  
\usepackage{graphicx}
\usepackage{txfonts}
\def\about{$\sim$}
\def\Fnu{$\nu$ F$_{\nu}$}
\def\msol{M$_{\odot}$\ }

\begin{document}
   \title{INTEGRAL observations of the blazar Mrk 421 in outburst}

   \subtitle{Results of a multi-wavelength campaign}

   \author{G. G. Lichti \inst{1} \and E.~Bottacini\inst{1} \and M.~Ajello\inst{1}
\and P.~Charlot\inst{2} \and W.~Collmar\inst{1}
\and A.~Falcone\inst{3} \and D.~Horan\inst{4} \and S.~Huber\inst{5} \and \\ A.~von~Kienlin\inst{1}
\and A.~L\"ahteenm\"aki\inst{6} \and E. Lindfors\inst{6}${^,}$\inst{7} \and D.~Morris\inst{3} \and K.~Nilsson\inst{7}
\and D.~Petry\inst{1} \and M.~R\"uger\inst{5} \and A.~Sillanp\"a\"a\inst{7} \and F.~Spanier\inst{5}
\and M.~Tornikoski\inst{6} }

   \offprints{G. G. Lichti \\ e-mail: grl@mpe.mpg.de}

   \institute{
              Max-Planck-Institut f\"ur extraterrestrische Physik, Giessenbachstrasse, 85748 Garching, Germany
	\and
              Laboratoire d'Astrophysique de Bordeaux, Universit\'e Bordeaux 1, CNRS, BP 89, 33270 Floirac, France
	\and
	      Pennsylvania State University, Department of Astronomy \& Astrophysics, University Park, PA 16802 USA
	\and
	      High-Energy Physics Division, Argonne National Laboratory, Argonne, USA
	\and	
	      Universit\"at W\"urzburg, Lehrstuhl f\"ur Astronomie, Am Hubland, 97074 W\"urzburg, Germany
	\and
	      Mets\"ahovi Radio Observatory, Helsinki University of Technology TKK, Kylm\"al\"antie 114, FIN-02540 Kylm\"al\"a, Finland
	\and
	      Tuorla Observatory, V\"ais\"al\"antie 20, FIN-21500 Piikki\"o, Finland
             }

   \date{Received 6 December 2007; accepted 6 March 2008}

 
   \abstract{If one wants to understand the physics of blazars, better
   simultaneous observations are important at all wavelengths, so it
   was fortunate that a ToO observation of the TeV-emitting blazar
   Mrk~421 with INTEGRAL could be
   triggered in June 2006 by an increase in the RXTE count rate to 
   more than 30 mCrab. The source was then observed with all INTEGRAL 
   instruments, with the exception of the spectrometer SPI, for a total 
   exposure of 829 ks. During this time several outbursts were observed 
   by IBIS and JEM-X. Multiwavelength observations were immediately 
   triggered, and the source was observed at radio, optical, and X-ray 
   wavelengths up to TeV energies.}
   {The data obtained during these observations were analysed with
   respect to time variability, time lags, correlated variability,
   and spectral evolution and then compiled in a \Fnu spectrum.}
   {The observations of the different instruments/telescopes were
   analysed with the usual correlation and time-analysis methods.
   The spectral analysis of the X-ray data was performed with XSPEC.}
   {Four strong flares at X-rays were observed that were not seen
   at other wavelengths (partially because of missing data).
   From the fastest rise in the X-rays, an upper limit could be derived to the extension of the emission region.
   A time lag between high-energy and low-energy X-rays was observed, which
   allowed an estimation of the magnetic-field strength. The spectral analysis of the X-rays
   revealed a slight spectral hardening of the low-energy (3-\about43 keV) spectral index.
   The hardness-ratio analysis of the Swift-XRT (0.2 - 10 keV) data indicated a small correlation with the
   intensity; i. e., a hard-to-soft evolution was observed.
   At the energies of IBIS/ISGRI (20 - 150 keV), such correlations are less obvious. A multiwavelength
   spectrum was composed and the X-ray and bolometric luminosities calculated.}
   {The observed flaring activity of Mrk~421 is mainly visible at X-rays. It is found that
   the spectral change with intensity is small. But at least one flare
   showed a completely different spectral behaviour than the other flares, so one can
   conclude that each blob of relativistic-moving electrons has its own individual physical
   environment that leads to different emission characteristics. From a fit of a leptonic
   emission model to the data, one finds that the observed variability may be due to a varying
   efficiency of particle acceleration.}


\maketitle

%

\section{Introduction}

   In 1992 the first AGN, Mrk 421, was detected at energies $>$500~GeV (Punch et al. \cite{Punch}). Since then, 20
detections of AGNs have been reported at TeV energies (Blazejowski et al. \cite{Blaz}; Padovani et al. \cite{Padovani};
\cite{Wagner}). Seventeen of these objects belong to the HBL Lac type of AGNs. They are radio-loud sources
with the radio emission originating mainly in a core region and they are characterised by a high polarization
at radio and optical wavelengths and a strong variability at all wavelengths. The spectral characteristics point
to non-thermal emission processes that presumably take place in a narrow relativistic jet pointed at a small
angle to the line of sight. The spectral energy-density distribution of these sources shows two smooth broadband
emission components: a first one that reaches a broad peak in the IR to X-ray region and a second one at GeV to TeV energies.
Both emission components are clearly separated (see Figure~3 of \cite{GhiMaDo}).

It is believed by many researchers
that both components are generated by the same leptonic population (electrons and/or positrons) moving at relativistic
speed in the jet (Blandford and Rees \cite{Blandford}), which creates the low-energy photons via incoherent synchrotron
radiation and the high-energy gamma-rays via inverse-Compton scattering of soft photons. This common origin would
explain the similarity of the two components (Ghisellini and Maraschi \cite{Ghisell}); hence, a similar temporal
evolution is expected for both components if this scenario is correct. Such time correlations were indeed observed
by observations at X-ray and TeV energies (Buckley et al. \cite{Buckley}; Catanese et al. \cite{Catanese}; \cite{Fossati07}),
and they even extend to the optical and radio wavelengths  (Katarzynski et al. \cite{Katar03}). This idea is
further supported by the observed polarization at radio wavelengths (\cite{Piner}). However, there are some exceptions
to these clear correlations, such as the orphan TeV flares reported by Blazejowski et al. (\cite{Blaz}) and
\cite{Krawczynski1}.
Unfortunately, measurements between 10 keV and 1 GeV that would allow us to further constrain this common origin have been
scarce. Until now only few measurements between 10~keV and 1~GeV exist. COMPTEL has detected Mrk 421 in the 10-30 MeV
range with 3.2$\sigma$ (Collmar et al. \cite{Collmar}). EGRET, however, has detected this source many times above 100 MeV
(Hartman et al. \cite{Hartman}).

Since the high-energy emission is not as well understood, many different models try to explain this emission component.
In one scenario the low-energy synchrotron photons are boosted to high energies by the same electron population that
creates the synchrotron photons [the synchrotron self-Compton (SSC) models (K\"onigl \cite{Konigl}, \cite{Mar} 1999,
Bloom \& Marscher \cite{Bloom})]. Although this process must be at work in all blazars and if the synchrotron-emission
hypothesis is correct, it may not be the dominant one. In another model the seed photons for the inverse-Compton effect
enter from outside the jet region, e. g. from the accretion disc or from clouds surrounding the jet (Dermer et al.
\cite{Dermer}; Sikora et al. \cite{Sikora}). Apart from these lepton models, the so-called hadron models were proposed
in which the high-energy $\gamma$-rays are produced by proton-initiated cascades (Mannheim \cite{Mannh},
M\"ucke \& Protheroe \cite{Mucke}) and/or proto-synchrotron emission (\cite{Aharonian05}; \cite{Muecke}).

The homogeneous SSC model makes very definite predictions about
the correlated behaviour of the high-energy end of both the synchrotron
and SSC components: simultaneous variability of photons and well-defined
correlated spectral changes in the medium/hard X-rays and TeV bands.
But these models have problems with the bulk-Lorentz factor statistics
(Henri \& Saug\'e \cite{Henri}) and it seems to be very important to measure a
detailed variability pattern from X-rays up to TeV $\gamma$-rays. Several
alternatives have been proposed, and each needs to be quantified in
more detail. However, in general one can say that in the leptonic
models synchrotron radiation is ``primary'' and inverse Compton radiation
``secondary'', while for at least one hadronic model it is the opposite:
the $\gamma$-rays are produced by the ``primary'' protons through cascades,
while the X-rays have a synchrotron origin in the ``primary'' electrons.
In this case the relation between the two components can be looser
than for lepton models. Also, in this case the ``cascade" spectra
are quite soft (lots of soft $\gamma$-rays).

To investigate all these theoretical ideas and models, an INTEGRAL ToO proposal was submitted to observe the
blazar Mrk~421 when it becomes active. Mrk~421 has a redshift of 0.031, and with a distance of about 125 Mpc [for 
$H_0$ = 71 km/(s Mpc)] IS one of the closest and therefore brightest blazars. It hosts a supermassive
black hole with a mass of (2--8) $\cdot 10^8$ \msol in its centre (\cite{Barth}; \cite{Falomo}; \cite{Treves}). At high energies,
MRK~421 has been detected in the 10--30 MeV range by COMPTEL with 3.2$\sigma$ (Collmar et al. \cite{Collmar}) and many
times above 100 MeV by EGRET (Hartman et al. \cite{Hartman}) and in the TeV range by Cherenkov telescopes (\cite{Kerrick};
\cite{Aharonian} 1999; \cite{Sambruna}; \cite{Krennrich02}; \cite{Aharonian04} and Rebillot et al. \cite{Rebillot}). In June 2006, Mrk ~421
became active and our INTEGRAL proposal was activated. The proposal foresaw not only observations with the INTEGRAL
instruments but also observations with other telescopes at all wavelengths. Data at radio, optical, X-ray, and TeV wavelengths
were taken.However, the 
observations at TeV energies turned out to be sparse because the visibility period for the Cherenkov telescopes 
approached its end. It was thus impossible to investigate possible correlations or time lags between the synchrotron 
and inverse-Compton emission as we wished. We therefore had to concentrate our analysis mainly on the INTEGRAL data
itself and on the multi-wavelength spectrum. Preliminary results of this analysis have been already published by \cite{Lichti}.

\section{OBSERVATIONS}

In April 2006 the blazar Mrk 421 increased its intensity to a level
$>$30 mCrab as measured by the all-sky monitor (ASM) of RXTE (the quiescent intensity
fluctuates strongly around an average value of 10 -- 15 mCrab). It remained
at this level until September 2006. This triggered an INTEGRAL observation
and correlated multiwavelength observations in the radio, optical, X-ray and TeV ranges.   
The details of the various observations are given in this section.

\subsection{INTEGRAL observations}

On June 14, 2006 (MJD 53900) at 09:53:16.8 hours UT,
an observation of this source was triggered with INTEGRAL. The observation
lasted about nine days and ended on June 26, 2006 (MJD 53911) at 02:03:55.4 hours UT.
In total the source was observed for 829 ks with the instruments IBIS,
JEM-X, and OMC in pointing mode (the spectrometer SPI was disabled because of annealing its Ge detectors).

The data of the three operating instruments were analysed with the INTEGRAL Off-line Scientific
Analysis (OSA) software version 5.1 using the latest response matrices available for that software.
The collected data were screened by computing the median count rate for each science window,
and it was then compared with each of the other science windows and their distributions.
Those science windows showing rates higher than 10 standard deviations from the median count rate were
checked again. By this screening effort, three suspicious science windows were detected. A closer look
revealed that these science windows belonged to times when INTEGRAL was close to the radiation belts
and that the high counting rates were due to the trapped charged particles. After this screening
process, 230 science windows remained and were considered for further analysis.

\begin{figure*}
   \centering
    \includegraphics[width=6cm]{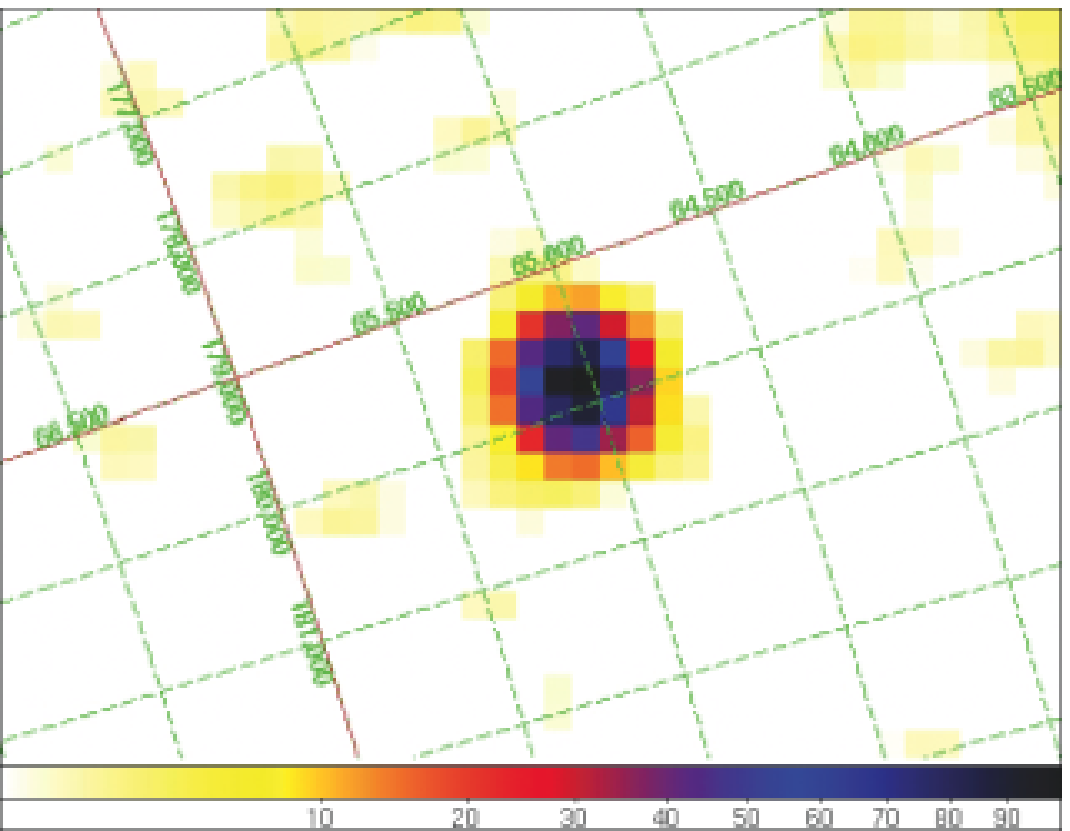}
    \includegraphics[width=6cm]{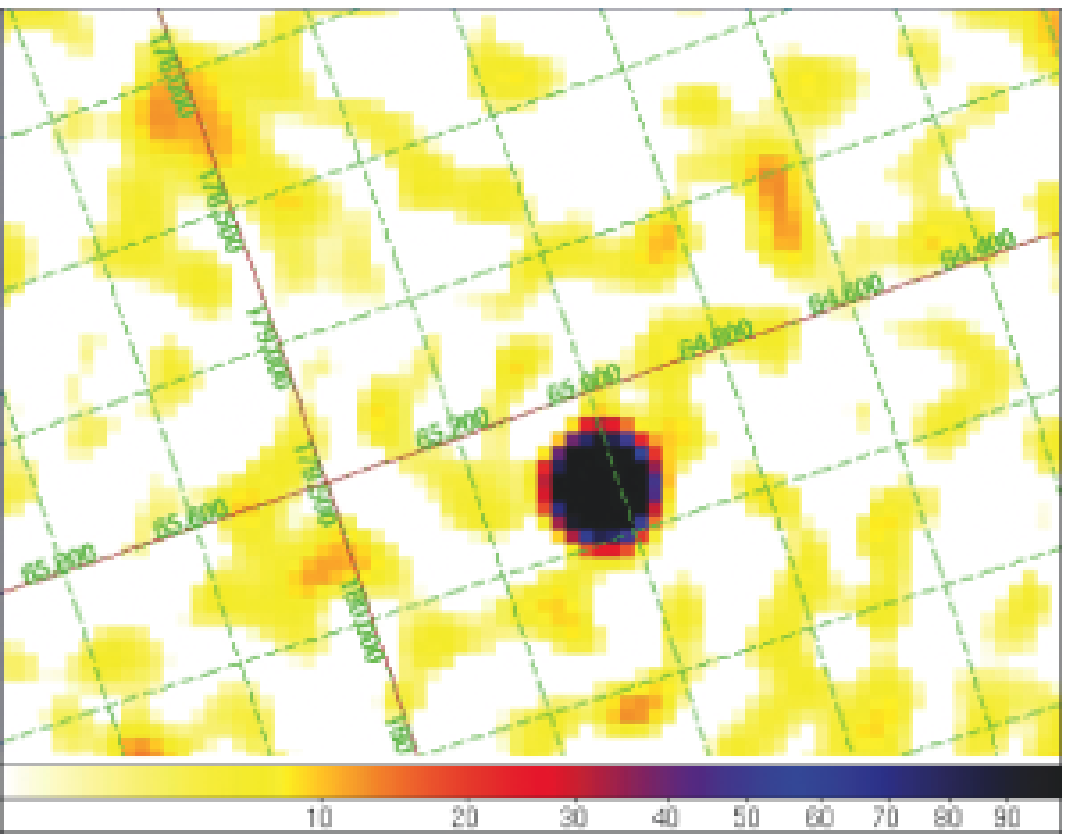}
    \includegraphics[width=6cm]{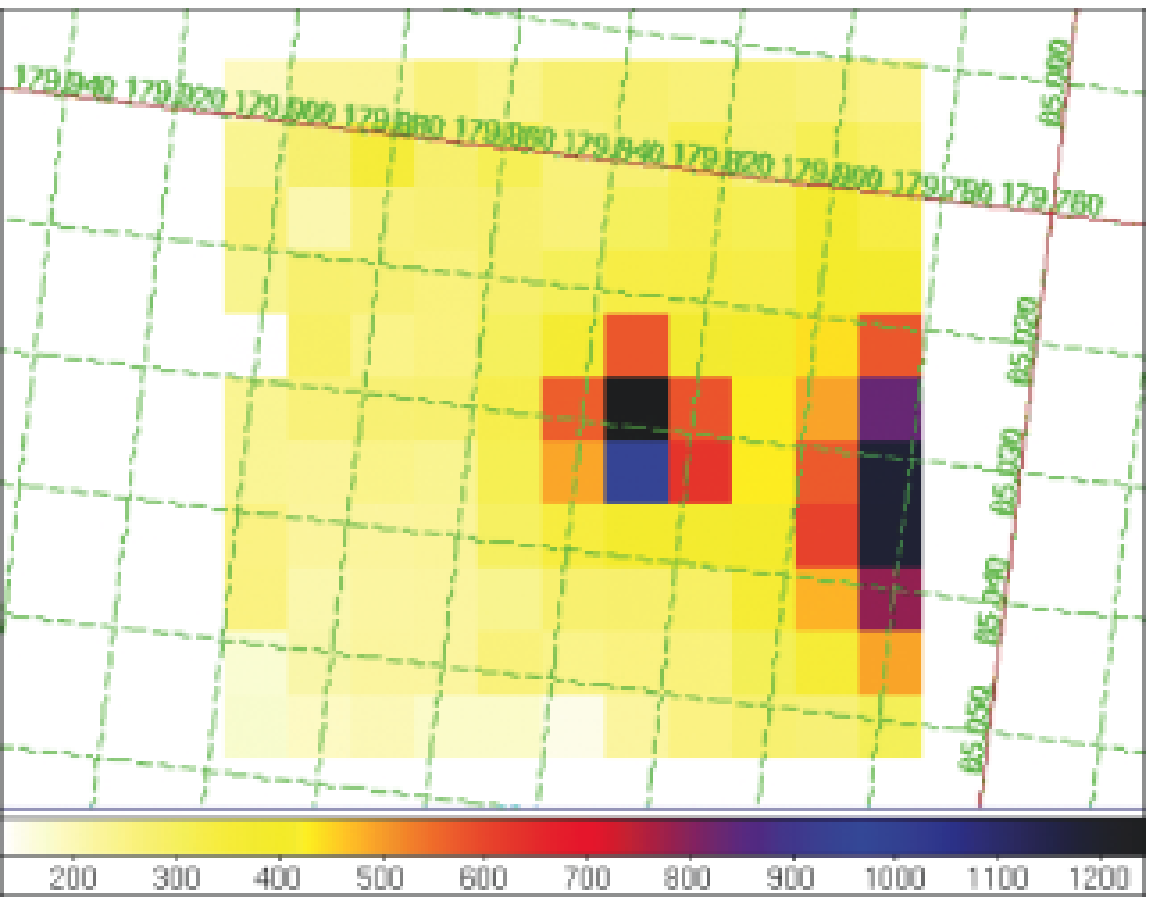}
   \caption{IBIS/ISGRI, JEM-X, and OMC skymaps (from left to right) of the region around Mrk 421 for 20-40 and 3-4.5 keV
and at visible wavelengths, respectively, in galactic coordinates. In the OMC map, the bright star
51 Uma (V = 6), which is clearly separated from Mrk 421 (distance $\sim$2 arcmin), is seen at the edge of the field of view.}
   \label{skymaps}
 \end{figure*}

Mrk 421 was clearly detected by all three INTEGRAL instruments. In Figure~\ref{skymaps} IBIS/ISGRI,
JEM-X, and OMC skymaps in galactic coordinates are shown. In all three maps the source is detected at a high significance level
[by IBIS in the energy interval 20--40 keV with 160$\sigma$, by JEM-X in the energy interval 3--4.5 keV with
720$\sigma$, and by OMC (limiting magnitude $\approx$18) with m$_v \approx$ 12.87)]. It should be mentioned that in the field of view
of IBIS another AGN, NGC 4151, was observeded, which was detected in the energy intervals 20-50 keV, 50-100 keV,
and 100-150 keV with 34.7$\sigma$, 15.1$\sigma$, and 5.4$\sigma$, respectively.

The magnitudes m of OMC were converted into differential flux values with the following formula
(obtained from A. D. Garau of the OMC team)

\begin{eqnarray}
S[{erg \over cm^2 s \AA}] & = & 3.64 \cdot 10^{-9} \cdot 10^{-0.4 m} \\
\Delta S & = & 0.921 \cdot S \cdot \Delta m.
\end{eqnarray}

\noindent The calibration of (1) was obtained from the flux of a star with m = 0 at 5500 $\AA$ from
\cite{Wamsteker}. The contribution of the host galaxy was estimated from the colour of a typical
elliptical galaxy (V--R = 0.6) and its R-band flux of 16.5 mJy to be 9.5 mJy. This host-galaxy flux
was taken into account when calculating the energy-density flux. A correction for a possible
contribution from the bright (V = 6) close-by ($\sim$ 2 arcmin away) star 51 Uma (HR 4309) is not
necessary because of the sharp point-spread function of the OMC, which drops practically to zero
at a distance of 2 arcmin (attenuation factor $<10^{-23}$).

Immediately after the beginning of the observation with INTEGRAL, the other
observers participating in this multiwavelength campaign were alerted, and
they commenced with their observations as allowed by the observational
constraints.

\subsection{VLBA observations}

Following the notification of the INTEGRAL observations, a request
to trigger our previously-approved target of opportunity VLBA
observations was sent to NRAO on June 13, 2006. After preparation of
the observing schedule, the project entered the queue for dynamical
scheduling at the VLBA and actually got scheduled on June 25. The
observations started at 17:00~UT on that day and lasted for
14~hours, with scans alternated on six different frequency bands
(2.3, 5.0, 8.4, 15.4, 22.2, and 43.2~GHz). The total observing time
on each frequency band was roughly 1~hour at 2.3, 5 and 8.4~GHz,
2~hours at 15~GHz, 3~hours at 22~GHz, and 6~hours at 43~GHz.
Additional scans on calibrator sources were also inserted into the
schedule for proper amplitude and polarization calibration.

After correlation at the Array Operation Center in Socorro, the data
were calibrated and corrected for residual delay and delay rate
using the NRAO Astronomical Image Processing System (AIPS) and
subsequently imaged using the Caltech VLBI imaging software DIFMAP.
The flux density was then integrated over each image to produce the
total flux densities used in the present study of the spectral
energy distribution of Mrk~421. The values of the flux density
derived so far in this way are 0.283~Jy at 5~GHz and 0.273~Jy at
8.4~GHz. Presentation of
the VLBI images is deferred to a future paper, which will present and
discuss the results of the VLBI data analysis in more details.

\subsection{Observations with the Mets\"ahovi radiotelescope}

The 37~GHz observations were made with the 13.7 m diameter Mets\"ahovi radio telescope, which is a
radome-enclosed paraboloid antenna situated in Finland (24 23' 38''E, +60 13' 05''). The measurements
were made with a  1 GHz-band dual-beam receiver centred at 36.8 GHz. The HEMPT (high-electron
mobility pseudomorphic transistor) front end operates at room temperature. The observations are ON--ON
observations, alternating the source and the sky in each feed horn. A typical integration time to
obtain one flux-density data point is 1200--1400~s. The detection limit of our telescope at 
37~GHz is approximately 0.2~Jy under optimal conditions. Data points with a signal-to-noise ratio
$<$~4 were handled as non-detections.

The flux-density scale was set by observations of the bright compact HII region DR~21.
Sources 3C~84 and 3C~274 were used as secondary calibrators.
A detailed description on the data reduction and analysis is given in Ter\"asranta et al.
\cite{Terasranta}. The error estimate in the flux density includes the contribution from the
measurement rms and the uncertainty of the absolute calibration.

The observations of Mrk 421 during the INTEGRAL campaign were made in varying weather conditions.
The first measurements were non-detections, but shortly the flux level increased and several
detections were achieved on days 53902 and 53903. The rest of the campaign was lost due to changeable
weather and the end of the telescope time allocation.

The sparse radio data do not allow any conclusion by themselves. For the Mets\"ahovi observations Mark~421 stays
basically on the verge of the detection limit and it never gets very bright. This is typical of the
TeV blazars observed at Mets\"ahovi. However, this kind of  behaviour is different from other GeV-peaking
gamma-ray blazars, and may reflect the different mechanism for the production of gamma rays in Mrk~421
and the other TeV blazars.

\subsection{KVA observations}

Mrk~421 was observed in the R-band (centred at 640~nm)
with the 35~cm KVA telescope on La
Palma, Canary Islands, during 8~nights on June 14-22, 2006. The camera
employed was an SBIG ST-8 with a gain factor of 2.3~e$^-$/ADU and and
readout noise of 14~e$^-$. Each night 4--10 exposures of 180~s were
made. The images were processed in a standard way (bias subtraction,
dark subtraction and flat-fielding) and the raw counts of Mrk~421 and
stars 1-3 in Villata et al. \cite{Villata} were integrated within a 15~arcsec
diameter aperture. The magnitude of Mrk 421 was determined by
comparing its counts to star~1, for which R = 14.04 from Villata et
al. \cite{Villata} was assumed. The use of differential mode effectively
eliminates the influence of varying transparency due to clouds,
etc., and accurate fluxes can even be obtained under varying observing
conditions. The flux of Mrk~421 was then computed from

\begin{equation}
F[Jy] = 3080.0\cdot10^{-0.4 \cdot R}\ ,
\end{equation}

\noindent
where R is the R-band magnitude (with an effective wavelength of 6400~$\AA$).
The normalization constant was taken from Table IV of \cite{Bessell}.
The host galaxy flux with the 15~arcsec diameter aperture is 8.0$\pm$0.5 mJy
(Nilsson et al. \cite{Nilsson}), and this flux was subtracted from the observed
fluxes before proceeding with the analysis.

\subsection{RXTE observations}

The data of the all-sky monitor (ASM) of RXTE (for a description see Levine et al. \cite{Levine}) monitors
the X-ray sky more or less continuously in the energy range 1.5--12~keV.
Its data are publicly available. The data for Mrk~421 were
downloaded and the fifth lightcurve at the top of Figure~\ref{all-LC-new} was produced. The RXTE counting rates were translated
into energy fluxes by using the Crab-nebula flux from Kirsch et al. \cite{Kirsch}. These authors
derived an energy flux of $2.2 \cdot 10^{-8}$ erg/(cm$^2$ s) for the Crab for the energy interval
2--10~keV (i. e. the energy range of the ASM). This corresponds to an RXTE counting rate of 73 counts/s,
so 1 RXTE count/s corresponds to $3 \cdot 10^{-10}$ erg/(cm$^2$ s). With this value, the
RXTE-counting rates were converted to energy fluxes.

\subsection{Swift-XRT observations}

The X-Ray Telescope (XRT) instrument on the Swift Observatory (\cite{Burrows} and \cite{Gehrels})
is sensitive to X-rays in the 0.2--10 keV band.  These data were reduced using the latest HEAsoft tools
(version 6.1.0), including Swift software version~2.0 and the latest response (version~8) and
ancillary response files (created using xrtmkarf) available in CALDB at the time of analysis.
Data were screened with standard parameters, including the elimination of time periods when
the CCD temperature was warmer than -48$^\circ$ C.  Only WT (windowed timing) mode data was used
in this analysis due to the high rates of the active source, and only grades 0--2 were included.
Since the count rate stayed below $\approx$100 c/s, the WT mode data is free of significant
pile-up effects. These data were corrected for effects due to bad columns and bad pixels.
Source and background regions were both chosen in a way that avoids overlap with serendipitous
sources in the image. All analysis and fitting of XRT data was done in the 0.3 to 10 keV 
energy band. Due to the low hydrogen-column density towards Mrk 421 of 1.43 $\cdot$ 10$^{20}$ cm$^{-2}$,
the absorption is $<$0.5\% and was neglected. 
Finally the measured counting rates were transformed into energy fluxes
using the formula

\begin{equation}
F[{erg \over (cm^2 s)}] = (4 \pm 2.5) \cdot 10^{-11} \cdot R_{cts/s}.
\end{equation}

This conversion factor was calculated using a subset of these Swift-XRT 
Mrk 421 observations, which included both high- and low-state flux 
measurements. For each time period used, the rate was calculated using 
the analysis described above, and the spectral fits to these data 
allowed the flux to be calculated, thus resulting in a conversion 
factor. Since a single conversion factor was used, rather than many 
individual time-resolved spectral fits, the large error bars are 
required to represent the variation in this conversion factor due to 
spectral variability of the source.

\subsection{Swift-BAT observations}

The Burst-Alert Telescope (BAT, \cite{Barthelmy}) on board the Swift satellite mission
(\cite{Gehrels}) is a coded-mask telescope sensitive to the 15--200 keV energy range.
Thanks to the pointing strategy and to its wide field of view (FoV), BAT surveys \about80\% of
the sky every day. We thus looked in the archive for BAT observations that contained Mrk 421
in the FoV.  We selected all observations included in the time-span June 12--26, 2006. 
The BAT data were processed using the HEASOFT 6.2 package and according to the recipes presented in
http://swift.gsfc.nasa.gov/docs/swift/analysis/threads/bat\_thread s.html.
Spectra and lightcurves were corrected for off-axis variation of the rates
and residual background contamination as described in \cite{Ajello}.

\subsection{Whipple observations}

The Whipple Telescope is located at the Fred Lawrence Whipple Observatory,
at an elevation of 2300~m in Southern Arizona. It comprises a 10~m dish
on which 248 mirrors are mounted. These mirrors reflect the Cherenkov light
from extensive air showers onto a 379-element imaging camera at the focal plane 
of the telescope. The instrument is described in detail by \cite{Kildea}.

The Mrk~421 data presented here were taken in ``{\it{tracking}}'' mode at large zenith
angles on June 18, 19, and 21, 2006 (MJD 53904, 53905, and 53907).  The observing scans were
performed with the gamma-ray source at the centre of the field of view. No separate control
data were taken. Rather, the background rate of gamma-ray-like events was estimated from the
distribution of events passing all but the orientation image-selection cuts \cite{Horan}.
The gamma-ray rates are calculated in units of the flux from large zenith-angle observations
of the Crab Nebula, the standard candle in TeV astronomy, and are given in Table~\ref{table:A}.
They were converted into energy-flux values by integrating the Crab spectrum of \cite{Hillas}
from the threshold energies of 0.6 and 0.9 TeV to infinity and using the resulting flux values
as normalization for the time-averaged spectrum of Mrk 421 as given by \cite{Aharonian} (1999).
Under the assumption that the spectral shape is constant this normalization was adapted to
the measured flux values.

\begin{table}
\begin{minipage}[t]{\columnwidth}
\caption{The measured TeV intensities (in Crab units) of Mrk 421 as measured by the Whipple observatory.}       
\label{table:A}      
\centering                          
\renewcommand{\footnoterule}{}  
\begin{tabular}{c c c c c}        
\hline\hline                 
MJD\footnote{The integration time for each data point was 28 minutes. The centre of each time interval is given.}
    & flux value & error of & energy density & energy-den- \\    
    & (Crab units) & flux value & [erg/(cm$^2$ s)]& sity error\\ 
\hline                        
   53904.17 & 0.25 & 0.38 & $2.90 \cdot 10^{-11}$ & $4.46 \cdot 10^{-11}$ \\      
   53905.17 & 1.63 & 0.42 & $1.89 \cdot 10^{-10}$ & $6.31 \cdot 10^{-11}$ \\
   53907.20 & 0.39 & 0.28 & $3.71 \cdot 10^{-11}$ & $2.82 \cdot 10^{-11}$ \\ 
\hline                                   
\end{tabular}
\end{minipage}
\end{table}

\section{Results of the timing analysis}

It is known that blazars are time variable on all different time scales. This is especially
true for TeV blazars and thus also for Mrk 421. We have therefore investigated our data in this respect
and compare the data collected at different energies with each other and search for possible
correlations betweeen them. The results of this exercise are presented in this section.

\subsection{The Lightcurves}

\begin{figure*}
   \centering
   \includegraphics[width=18cm]{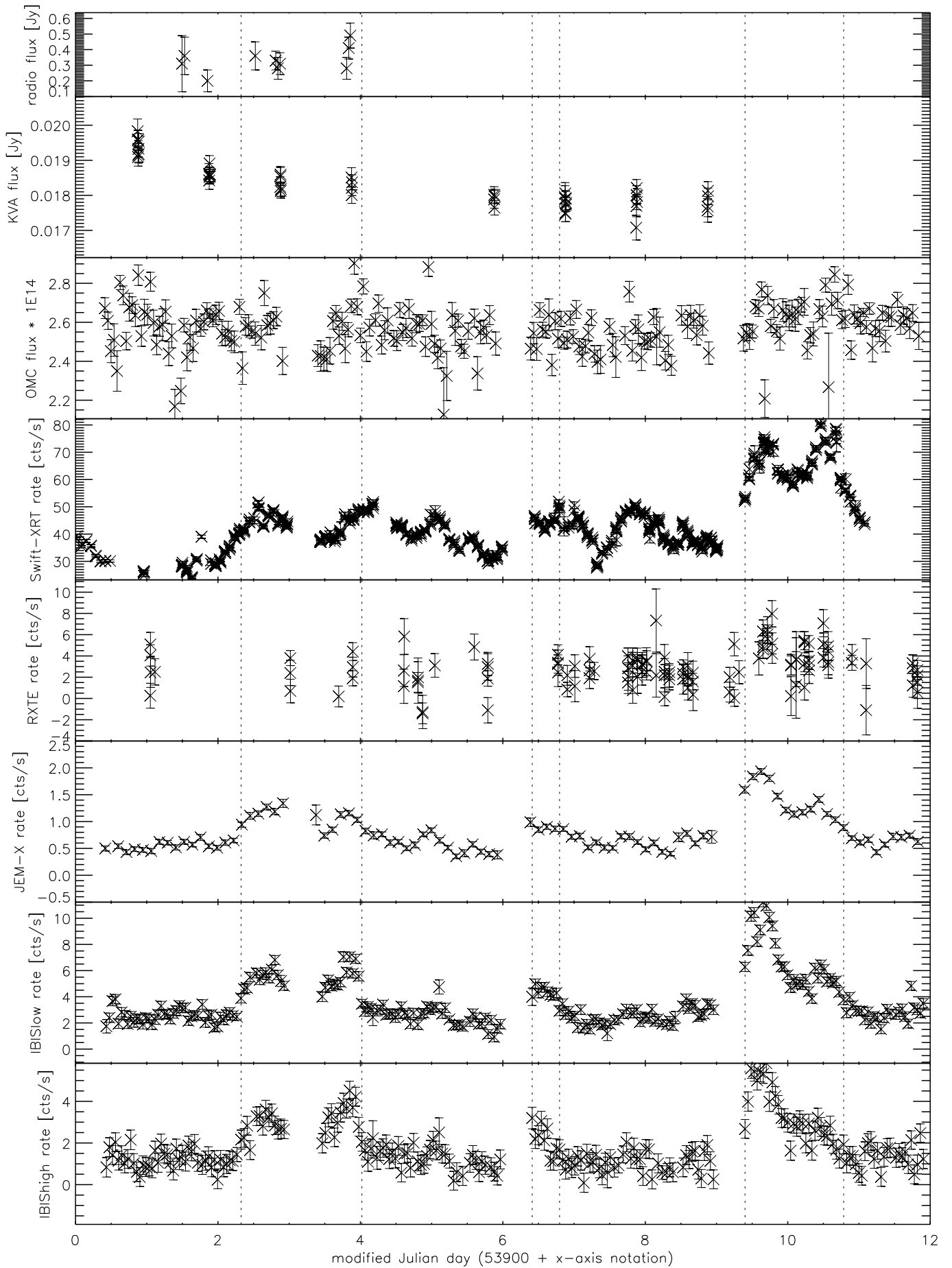}
   \caption{All lightcurves for Mrk~421 at radio, optical, X-ray, and gamma-ray wavelengths. The data of the OMC
[given in erg/(cm${^2}$ s \AA)] and KVA telescopes were not extinction-corrected, since the extinction E$_{B-V}$
in the direction of Mrk 421 is only 0.03. The time intervals for the active phases are indicated by the dotted lines.}
   \label{all-LC-new}
 \end{figure*}

\begin{figure}
   \centering
   \includegraphics[width=7.5cm]{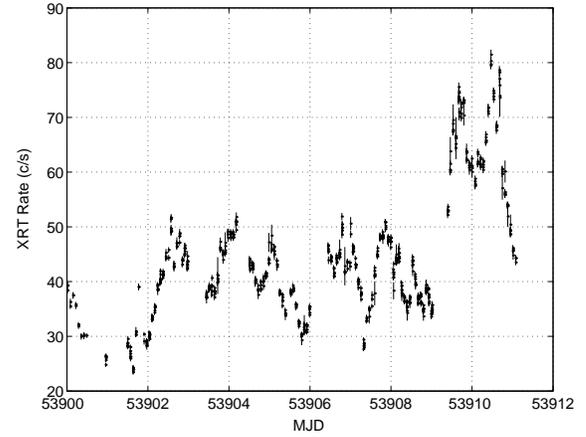}
   \caption{Swift-XRT lightcurve in the energy interval 0.3 - 10 keV. The dotted lines give only the coordinates, not the intervals
of the active phases as in Figures~\ref{JEMX-LC} to \ref{ISGRIhigh-LC}.}
   \label{XRT-LC}
 \end{figure}

\begin{figure}
   \centering
   \includegraphics[width=7.5cm]{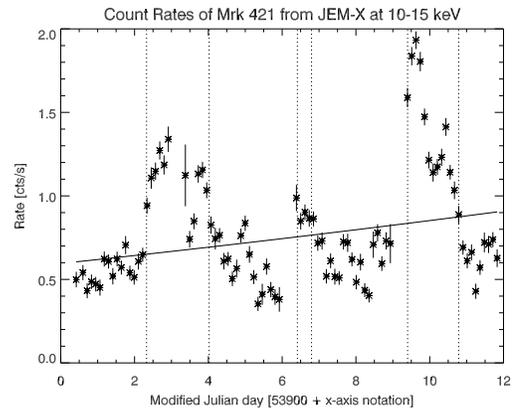}
   \caption{JEM-X lightcurve in the energy interval 10 - 15 keV. The time intervals for the
active phases are indicated by the dotted lines. A parabola was fitted to all counting rates to guide the eyes.}
   \label{JEMX-LC}
 \end{figure}

\begin{figure}
   \centering
   \includegraphics[width=9cm]{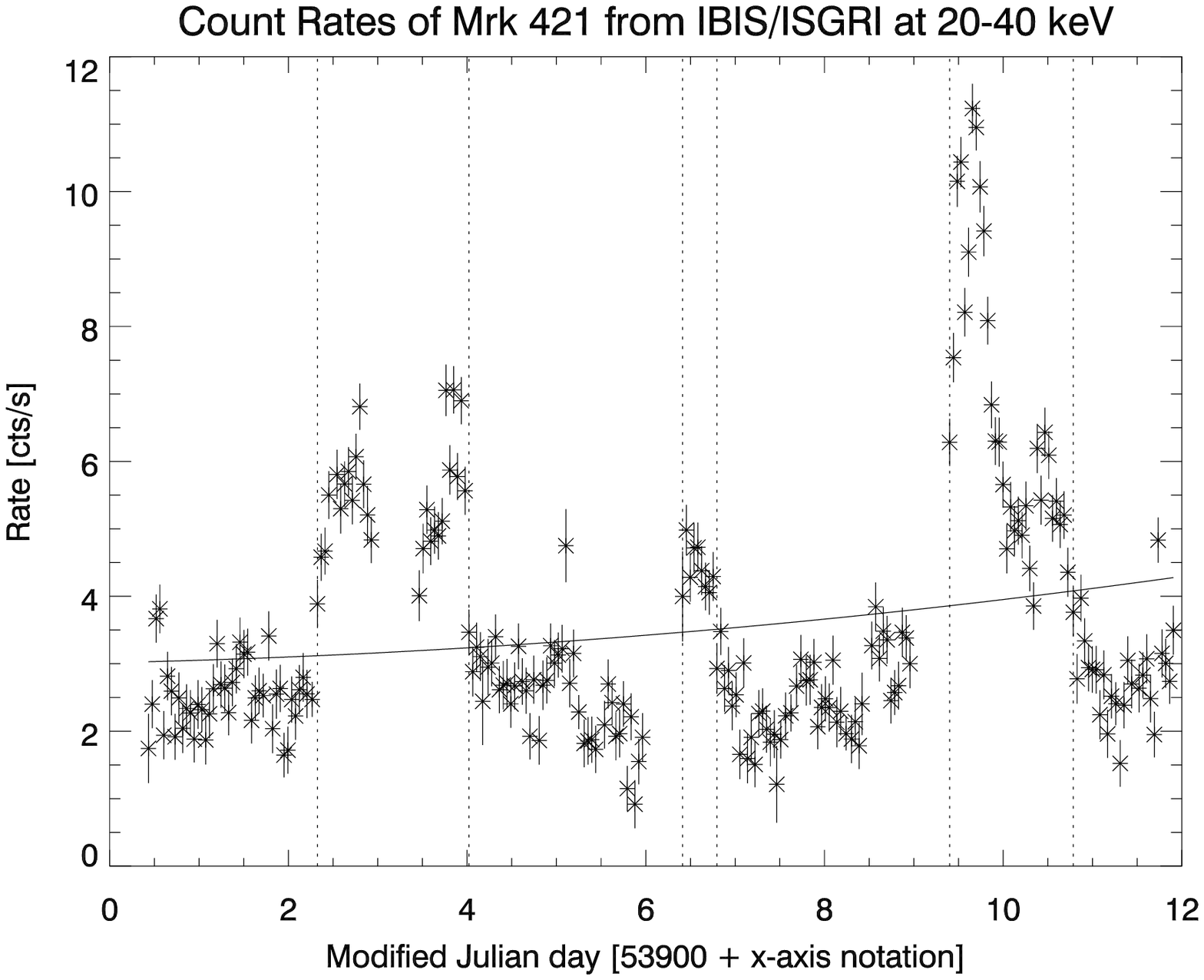}
   \caption{IBIS/ISGRI lightcurve in the energy interval 20-40 keV. The time intervals for the
active phases are indicated by the dotted lines. A parabola was fitted to all counting-rate values to guide the eyes.}
   \label{ISGRIlow-LC}
 \end{figure}

\begin{figure}
   \centering
   \includegraphics[width=9cm]{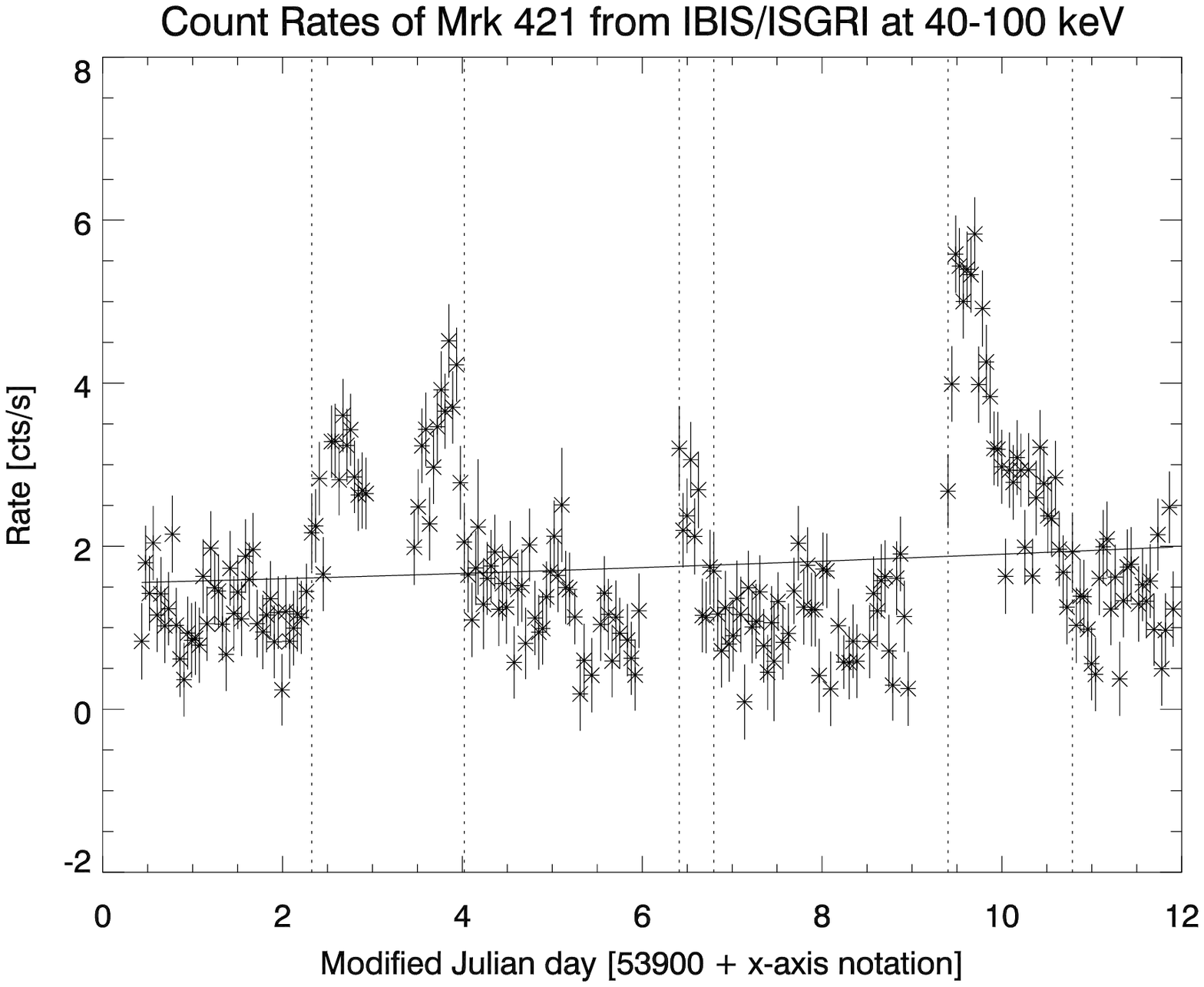}
   \caption{IBIS/ISGRI lightcurve in the energy interval 40-100 keV. The time intervals for the
active phases are indicated by the dotted lines. A parabola was fitted to all counting-rate values to guide the eyes.}
   \label{ISGRIhigh-LC}
 \end{figure}

\begin{figure}
   \centering
   \includegraphics[width=7.5cm]{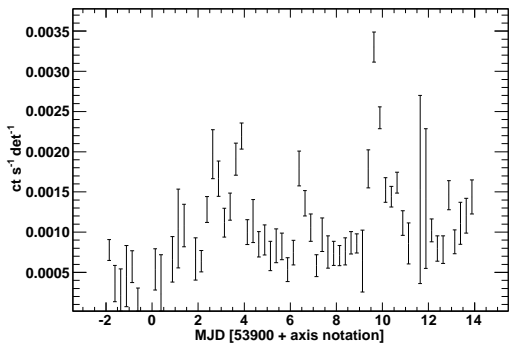}
   \caption{Swift-BAT lightcurve in the energy interval 15 - 200 keV.}
   \label{BAT-LC}
 \end{figure}

The lightcurves of the various observations are shown in Figures~\ref{all-LC-new} to \ref{BAT-LC}.
Whereas the sampling of the lightcurves of the three INTEGRAL instruments and the SWIFT-XRT telescope
is ample, this is not the case for the other lightcurves. This makes the search
for intensity correlations between the different wavebands difficult if
not even impossible. Another fact is striking when one looks at the lightcurves.
Although four strong flares are seen in the ISGRI lightcurve (and even more in the Swift-XRT
lightcurve of Figure~\ref{XRT-LC}), these flares are not visible at the other energies.
An inspection of the X-ray lightcurves reveals the following:

\begin{itemize}
  \item the strength and the shapes of the flares vary significantly;
  \item the shape of the flares is not symmetric (some flares show a sharp rise and a slow decay time,
        others a slow rise and a fast decay time);
  \item the duration of the flares is about 0.5 days;
  \item comparing Figures~\ref{XRT-LC} and \ref{JEMX-LC} one can see that the energy spectrum of
the flare around day 53910.5 is qualitatively softer than the spectrum of the other flares;
  \item from a linear fit to all the data, it becomes obvious that the total intensity is slightly
        increasing with time.
\end{itemize}

Especially interesting is that these flares are also not seen by the OMC. It is the common understanding
that in blazars, if viewed under small angles, the optical thermal emission of the accretion disk and
of the galaxy is outshined by the non-thermal beamed emission of the jet (Tavecchio \cite{Tavecchio}).
Since both the X-rays and the optical photons are produced by the same population of
relativistic electrons via synchrotron emission, one would expect similar (if not even identical) lightcurves,
if this scenario is correct. For Mrk~421, this happened to be the case in 2001 when a flare showing variability 
in the visible band coincident with that at radio wavelengths and at TeV energies was observed
(Katarzynski et al. \cite{Katar03}). That this is not observed in our data contradicts to this hypothesis
and asks for an explanation. Possible explanations will be discussed in the last section.

From the radio, KVA, OMC, and RXTE lightcurves of Figure~\ref{all-LC-new}, the average fluxes/magnitude/counting rates
were calculated. The weighted means are given in Table~\ref{table:x}. The ISGRI lightcurve of Figure~\ref{ISGRIlow-LC}
was used to define by eye the times when the source was in quiescent and active states. Three time intervals were
specified and their boundaries are given in Table~\ref{table:1}.

\begin{table}
\caption{Weighted means of the fluxes/magnitude/counting rates in Figure~\ref{all-LC-new}.}             
\label{table:x}      
\centering                          
\begin{tabular}{c c c}        
\hline\hline                 
instrument & average value & error \\    
\hline                        
   radio & 0.33 Jy & 0.026 Jy \\      
   KVA & 1.83 $\cdot$ 10$^{-2}$ Jy & 3 $\cdot$ 10$^{-4}$ Jy\\
   OMC & 2.579$\cdot$10$^{-14}$ erg/(cm${^2}$ s \AA) & 0.007 erg/(cm${^2}$ s \AA) \\
 RXTE & 2.86 cts/s & 0.3 cts/s \\ 
\hline                                   
\end{tabular}
\end{table}

\begin{table}
\caption[]{Time boundaries in modified Julian Date (days:hours:\\minutes:seconds) for the active state.}   
\label{table:1}      
\centering                          
\begin{tabular}{c c c}        
\hline\hline                 
interval number & start time & end time \\    
\hline                        
   1 & 53902:07:47:23 & 53904:00:29:59 \\      
   2 & 53906:09:53:24 & 53906:19:07:15 \\
   3 & 53909:09:36:14 & 53910:18:49:56 \\ 
\hline                                   
\end{tabular}
\end{table}

\subsection{Rise-time analysis}

The rise-time scale gives information about the extension of the emission region. From observations at
TeV energies it is known that Mrk 421 shows short-time variability with time scales of about one
day and shorter (Aharonian et al. \cite{Aharonian02}, \cite{Aharonian03}). Doubling times with time scales
of about one day or even down to 15-20 minutes were also reported by Cortina and Schweizer \cite{Cortina}
and
Krennrich et al. \cite{Krennrich}. In May 1996 two fast flares were observed (Gaidos et al. \cite{Gaidos}).
The first one had an intensity 10 times higher than the one of the Crab nebula (the highest flux
ever observed from this source at these energies) with a time scale of $<$1 day. The second one a week later
had an unprecedented time scale of \about 30 minutes indicating a very small emission region!

An inspection of the ISGRI ligtcurves reveals that the steepest ascent occurred in the third time interval.
Fitting an exponential rise-time law of the form $a \cdot e^{t \over t_0}$ to the data 
of the energy interval 40 - 100~keV one finds a value of 2.79 hours for $t_0$. In the energy interval
20 - 40~keV, the increase is somewhat slower (only 4.28 hours).
One can now calculate the size of the emission region with the formula

\begin{equation}
l \le {{c \cdot t_0 \cdot \delta} \over {1 + z}}
\end{equation}

\noindent
with $c$ the speed of light, $\delta$ the Doppler factor (= 12; from Kino et al. \cite{Kino}) and $z$
the redshift. Inserting the value from above one means that the emission region must be $<$ 234 AU
(the Schwarzschild radius of Mrk 421 is between 4 and 16 AU and the radius of the last stable
orbit between 12 and 47 AU). From the shortest time
scale measured at TeV energies an extension was derived of the emission region of $\le$ 2.5 AU
(Aharonian et al. \cite{Aharonian02}). From pure high-resolution radio-interferometric observations
Charlot et al. \cite{Charlot} derived an upper limit on the high-energy emission region of 0.1 pc
($\simeq$ 20000 AU). Thus a dependence of the size of the emission region on energy seems to exist:
the higher the energy, the smaller the emission region!

\subsection{Variability analysis}

The fractional variability V (see appendix in Fossati et al. \cite{Fossati} for the definition) was
calculated for all lightcurves. The result is shown in Table~\ref{table:var}. The same behaviour as
already found by previous authors (e. g. Fossati et al. \cite{Fossati}, Sembay et al.
\cite{Sembay}) is confirmed by our observations: the fractional variability increases with
energy. This behaviour seems to be common in blazars (Ulrich et al. \cite{Ulrich}).
If we fit a power law to the data we get V $\propto$ E$^{0.28}$. This is in good agreement
with the laws found by Fossati et al. (\cite{Fossati}) (V $\propto$ E$^{0.25}$) and by Giebels et al.
(\cite{Giebels}) (V~$\propto$~E$^{0.24}$). This reflects that the higher energies are
emitted from regions closer to the central black hole, which are smaller in extension than regions
farther away. Since the emission from small regions can vary faster than the
ones from larger regions the observed trend can be understood qualitatively.

\begin{table}
\caption{The fractional variability measure V for the different lightcurves.}        
\label{table:var}      
\centering   
\begin{tabular}{c c c c}        
\hline\hline                 
instrument & low-energy & high-energy & V \\    
           & bound [eV] & bound [eV] &    \\
\hline                        
Mets\"ahovi & - & - & undefined \\      
KVA         & 1.85 & 2.04 & 0.0314 \\
OMC         & 2.14 & 2.39 & 0.0393 \\
Swift-XRT   & 300  & 10000 & 0.266  \\
RXTE        & 2000 & 10000 & 0.4258 \\
JEM-X       & 10000 & 15000 & 0.4266 \\
IBIS/ISGRI  & 20000 & 40000 & 0.5146 \\
IBIS/ISGRI  & 40000 & 100000  & 0.5723 \\
\hline                                   
\end{tabular}
\end{table}

\subsection{Time-lag analysis}

\begin{figure}
   \centering
   \includegraphics[width=9cm]{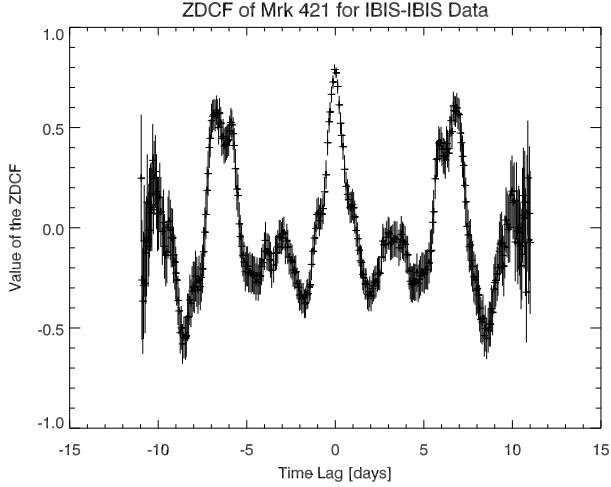}
   \caption{The Z-discrete correlation function of the IBIS data.}
   \label{ZDCF}
\end{figure}

The time lag between the two ISGRI lightcurves of the energy intervals 20-40 keV and 40-100 keV was
calculated with the Z-transformed discrete-correlation function (ZDCF; Edelson \& Krolik \cite{Edelson},
Tal \cite{Tal}). The value of the ZDCF as a function of the time lag $\tau$ is shown in Figure~\ref{ZDCF}.
Five maxima with ZDF values around 0.5 are seen, two at $\tau \approx \pm$ 3.4 days, two at $\tau \approx
\pm$ 7 days, and one at $\tau \approx$ 0 day. The shifts at $\pm$ 3.4 days and $\pm$ 7 days correspond to
the time difference between the flares seen in Figures~\ref{ISGRIlow-LC} and \ref{ISGRIhigh-LC} and thus
do not reveal new information. Important, however, is the question whether the
central peak has its maximum really at $\tau \approx$ 0 days. To find this out, a Gaussian and
a linear background were fitted to the data from $\tau$ = $-1$ day to $\tau$ = $+1$ day. The data can be
described well with a Gaussian ($\chi^2_{red}$ = 0.114). The fit to the data reveals that, with a
significance of 1.8$\sigma$, the maximum is not at a time lag of zero, but at
$\tau$ = ($-40.8 \pm 23.3$) minutes. The negative sign indicates that the variations at lower energies
lag behind those at the high energies.

The same time-lag analysis was performed for the IBIS (20-40~keV) / JEM-X (10-15~keV) and
the JEM-X (10-15~keV) / Swift-XRT (0.3-10~keV) lightcurves. In the first case the time lag is consistent
with a lag of zero minutes (actually $-4.25 \pm 30.4$~minutes), and in the second case the time lag is
positive and has a value of ($102.6 \pm 59$) minutes. Similar time lags (35 - 47 minutes) were
found by Fossati et al. (\cite{Fossati}) for the energy ranges 0.1--1.5~keV and 3.5--10~keV
using BeppoSAX data and by \cite{Takahashi} as a function of energy for the energy
range 0.5--7.5~keV using ASCA data. Especially \cite{Takahashi} found that the time lags
decrease with increasing energy, consistent with our results. But their time lags were all negative,
while we also find positive timelags. Positive time lags were also measured by \cite{Ravasio} and by
\cite{Brinkmann} in the energy range 0.6--10 keV using XMM data. But the statistical significance of
our time lags is marginal, so one can question their reality. The importance of this question
and the ample number of lightcurves at different energies measured during our multiwavelength campaign
would deserve a more detailed analysis, which is, however, not the scope of this paper.

The time lags must be produced by energy-dependent mechanisms, for instance, particle cooling, and acceleration.
The higher-energy particles cool faster and accelerate slower than the lower-energy particles, therefore
the negative timelag can be used to estimate the magnetic-field strength from the measured negative time lag.
The theory of this process was
developed by Kazanas et al. \cite{Kazanas}. Based on the results of this theory Chiapetti et al.
(\cite{Chiapetti}) derived the equation (\ref{B-field}) which provides an estimate of the magnetic field B:

\begin{equation}
B = 300 \cdot {\bigl({1+z \over \nu_1 \cdot \delta} \bigr)}^{1 \over 3} \cdot \Biggl({{1 - \sqrt{\nu_1 \over \nu_0}} \over \tau}\Biggr)^{2 \over 3}
\approx {\bigl({300^3 \over \nu_1 \cdot \delta} \bigr)}^{1 \over 3} \cdot \tau^{-{2 \over 3}}
\label{B-field}
\end{equation}

\noindent where $\delta$ is the Doppler factor, z the redshift, and $\nu_0$ and $\nu_1$ are the frequencies of the
corresponding energy intervals (in units of 10$^{17}$ Hz) at which the time lag $\tau$ ($\tau$ in s)
has been measured.

For determining the frequencies $\nu_0$ and $\nu_1$ the mean energy for the energy intervals
given above were calculated using the power laws derived from the fit to the data (see Tables~\ref{table:4}
and \ref{table:5}). Inserting those, together with the other parameters from above, one gets a magnetic-field
strength of (0.08 $\pm$ 0.03) G.
Similar values (0.12~G) were found by Kino et al. (\cite{Kino}) using
the observables of Mrk 421 from Kataoka (\cite{Kataoka}), by Konopelko et al.
(\cite{Konopelko}) (0.1~G) and by Giebels et al. (\cite{Giebels}) (\about0.1~G).

But the value of the Doppler factor is highly uncertain. Rebillot et al. (\cite{Rebillot})
investigated two models with $\delta$ = 50 and 1000 (in their model a higher Doppler
factor corresponds to a smaller emission region). If we adopt these values, we find
values of (0.06 $\pm$ 0.02)~G and (0.02 $\pm$ 0.007)~G, respectively, for the magnetic-field strength B.
Values of the Doppler factor between 50 and 100 seem to be suggested by the data
(see Krawczynski et al. \cite{Krawczynski} and Mastichiadis and Kirk \cite{Kirk}). However,
such large Doppler factors are not observed with VLBI observations (e.g. \cite{Piner}), so this might
indicate a velocity structure in the jet (\cite{Georganopoulos}; \cite{Ghisellini}).

\subsection{Correlation analysis}

\begin{figure}
   \centering
   \vspace{-9pt}
   \includegraphics[width=10cm]{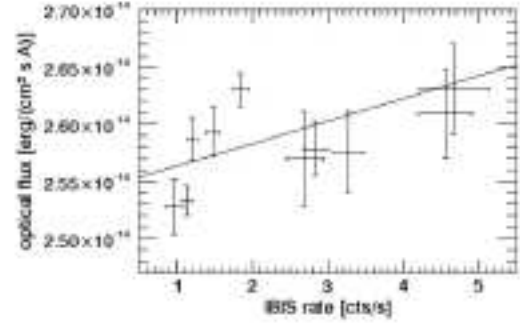}
   \vspace{-9pt}
   \caption{Intensity correlation between OMC and IBIS data. The count-rate range of the ISGRI
count rates was split into arbitrarily-chosen intervals given by the error bars.}
   \label{Figure4}
\end{figure}

We investigated whether the X-ray data of ISGRI correlate with the optical data of OMC. For this,
investigation time intervals were specified during which the background-subtracted ISGRI count rates of
Mrk 421 were lying in well-defined, but arbitrarily-chosen count-rate intervals.
The average ISGRI count rates and the average optical fluxes of OMC were calculated for these time
intervals. The result of this correlation analysis is shown in Figure~\ref{Figure4}. Although the
scatter of the results is
quite large, a slight increase in the optical flux with increasing ISGRI count rate is detected.
If one fits a straight line to the data, this dependence becomes more evident. The functional
dependence of the OMC flux on the ISGRI rate is

\begin{equation}
F_{OMC} = 2.5 \cdot 10^{-14} + 1.96 \cdot 10^{-16} \cdot R_{IBIS}$$.
\end{equation}

\section{Results of spectral analysis}

The results of a spectral analysis of the data of Mrk 421 at X-rays are presented in this section.
First, we investigated how the energy spectrum evolves when going from the quiescent into the active
state. Secondly,  we calculated hardness ratios and looked for possible correlations.
And third, we combined all data across the frequency range and constructed an energy-density
spectrum $\nu$ F$_\nu$.

\subsection{Spectral fits}

The data from JEM-X and IBIS were collected for the quiescent and active states according to the times
given in Table ~\ref{table:1}. Then different spectral models were fitted to the quiescent and active
fluxes using the functions available in XSPEC11. Since ISGRI and JEM-X were excellently cross-calibrated
by the two responsible teams, no cross-calibration factor had to be applied. 
The results of this fitting exercise are given in Tables~\ref{table:4} and \ref{table:5}.
In addition, a spectrum with a log-parabolic function was fitted to the
data (see Massaro et al. \cite{Massaro}). The results of these fits are shown in Table~\ref{table:6}.

\begin{table*}
\begin{minipage}[t]{\columnwidth}
\caption{Model fits to the JEM-X and ISGRI data of the quiescent state. The reference energy is 1 keV.}             
\label{table:4}      
\renewcommand{\footnoterule}{}  
\begin{tabular}{l c c c c c}        
\hline\hline                 
Model\footnote{The definitions of the mathematical forms of the used functions are contained in the XSPEC11 manual [with the exception of the joint PL which has the form 
${({E \over E_b})^{-\alpha} / (1 + ({E \over E_b})^{\beta-\alpha}})$].} & $\chi^2_{red}$ & normalization constant [keV$^{-1}$ cm$^{-2}$ s$^{-1}$] & $\alpha$\footnote{$\alpha$ is the spectral index at energies below E$_b$}
& $\beta$\footnote{$\beta$ is the spectral index at energies above E$_b$} & E$_b$ [keV]\footnote{E$_b$ is the break energy} \\    
\hline                        
 power law (PL) & 2.17 & (0.378 $\pm$ 0.004) & 2.318 $\pm$ 0.006 & - & - \\ 
 joint PL & 1.85 & (9.3 $\pm$ 3) $\cdot$ 10$^{-6}$ & 2.298 $\pm$ 0.007 & 6.5 $\pm$ 1.0 & 149 $\pm$ 11 \\
 {\bf broken PL} & {\bf1.83} & {\bf(0.365 $\pm$ 0.005)} & {\bf2.300 $\pm$ 0.006} & {\bf3.0 $\pm$ 0.2} & {\bf45 $\pm$ 4} \\
 Band model & 2.2 & (2.8 $\pm$ 5) $\cdot$ 10$^{-5}$ & 2.0 $\pm$ 0.2 & 2.32 $\pm$ 0.01 & 13 $\pm$ 2 \\
 PL with exponential cutoff & 2.17 & (0.355 $\pm$ 0.006) & 2.27 $\pm$ 0.01 & - & 311 $\pm$ 77 \\ 
\hline                                   
\end{tabular}
\end{minipage}
\end{table*}

\begin{table*}
\caption{Model fits to the JEM-X and ISGRI data of the active state. The reference energy is 1 keV.
The parameters and functions have similar meaning to the one in Table~\ref{table:4}.}             
\label{table:5}      
\begin{tabular}{l c c c c c}        
\hline\hline                 
Model & $\chi^2_{red}$ & normalization constant [keV$^{-1}$ cm$^{-2}$ s$^{-1}$] & $\alpha$ & $\beta$ & E$_b$ [keV] \\    
\hline                        
 power law (PL) & 3.67 & (0.530 $\pm$ 0.005) & 2.164 $\pm$ 0.004 & - & - \\ 
 joint PL & 1.85 & (9.9 $\pm$ 3) $\cdot$ 10$^{-6}$ & 2.298 $\pm$ 0.007 & 6.65 $\pm$ 1.1 & 145 $\pm$ 10 \\
 {\bf broken PL} & {\bf1.89} & {\bf(0.490 $\pm$ 0.005)} & {\bf2.12 $\pm$ 0.05} & {\bf2.90 $\pm$ 0.08} & {\bf41 $\pm$ 2} \\
 Band model & 3.85 & (9 $\pm$ 7) $\cdot$ 10$^{-5}$ & 1.81 $\pm$ 0.09 & 2.20 $\pm$ 0.02 & 23 $\pm$ 3 \\
 PL with exponential cutoff & 2.36 & (0.530 $\pm$ 0.005) & 2.04 $\pm$ 0.01 & - & 132.0 $\pm$ 0.3 \\ 
\hline                                   
\end{tabular}
\end{table*}

\begin{table*}
\caption{Results of fits to the JEM-X and ISGRI data with a log-parabolic function [= $K \cdot (E/E_1)^{-a-b \cdot log(E/E_1)}$]
for both states for a reference energy of E$_1$ = 40 keV.}  
\label{table:6}      
\begin{tabular}{l c c c c}        
\hline\hline                 
state of & $\chi^2_{red}$ & normalization constant & a & b \\    
 source &  & [keV$^{-1}$ cm$^{-2}$ s$^{-1}$] &  & \\
\hline                        
 quiescent & 2.23 & (7.21 $\pm$ 0.2) $\cdot$ 10$^{-5}$ & 2.35 $\pm$ 0.02 & (1.1 $\pm$ 0.6) $\cdot$ 10$^{-2}$ \\ 
 active & 2.91 & (1.77 $\pm$ 0.03) $\cdot$ 10$^{-4}$ & 2.28 $\pm$ 0.01 & (5.0 $\pm$ 0.5) $\cdot$ 10$^{-2}$ \\
\hline                                   
\end{tabular}
\end{table*}

It turned out that a broken power law of the form

\begin{eqnarray}
F(E) &=& A \cdot \Bigl({E \over 1 keV} \Bigr)^{-\alpha}\	for\ E \le E_b \\
F(E) &=& A \cdot E_{b}^{\beta - \alpha} \cdot \Bigl({E \over 1 keV} \Bigr)^{-\beta}\	for\ E > E_b     
\end{eqnarray}

\noindent
(with A the normalization constant and with $\alpha$ and $\beta$ the low- and high-energy spectral indices,
respectively) gave the best fit result in both cases. (The joint power law gave an equally good fit, but it leads to
an unrealistic high value of 6.5 for the high-energy power-law index $\beta$ and was therefore discarded.)

\begin{figure}
   \centering
   \includegraphics[width=6cm,angle=-90]{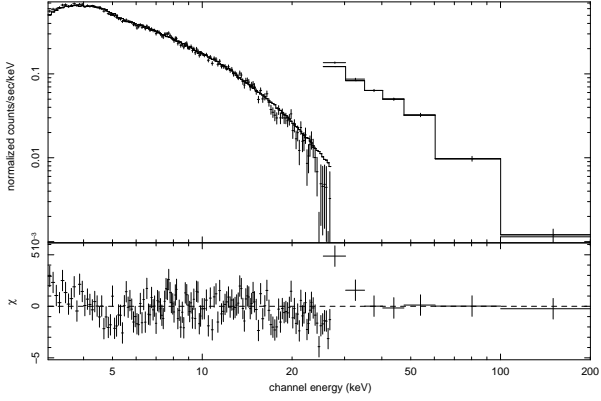}
   \vspace{-9pt}
   \caption{Result of the fit of a broken power law to the JEM-X and ISGRI data in the quiescent state. The top panel
shows the spectral data with the fit function, the bottom panel the residuals of the fit.}
   \label{quietspectrum}
\end{figure}

\begin{figure}
   \centering
   \includegraphics[width=6cm,angle=-90]{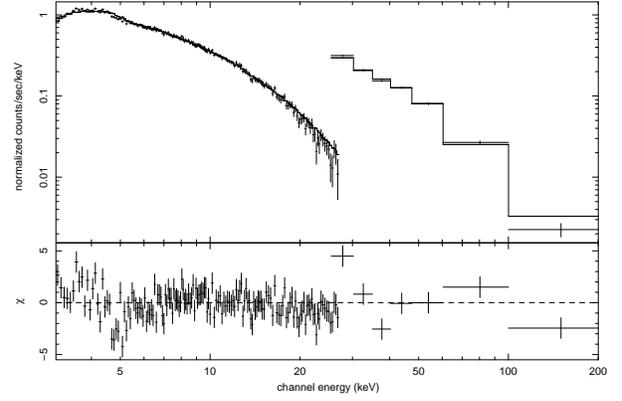}
   \vspace{-9pt}
   \caption{Result of the fit of a broken power law to the JEM-X and ISGRI data in the active state. The top panel
shows the spectral data with the fit function, the bottom panel shows the residuals of the fit.}
   \label{activespectrum}
\end{figure}

The results of the fits for the quiescent and active states are shown in Figures~\ref{quietspectrum}
and~\ref{activespectrum}, respectively.
With a reduced $\chi^2$ value of \about 1.9, the fits are not excellent, but reasonable.
The reason for the imperfect fit is that the fluxes of JEM-X and ISGRI in the overlapping energy
intervals differ as is seen from the residuals. The results of the spectral analysis can be summarised
as follows:

\begin{itemize}
 \item a broken power law fits the data best in both cases (quiescent and active)
 \item the break energy E$_b$ has a value of \about 43 keV
 \item the index $\alpha$ is larger than 2
 \item the spectral parameters do not change significantly with the state
  (although a slight spectral hardening of the low-energy index $\alpha$ is observed)
\end{itemize}

The third point indicates that we have only measured the declining part of the X-ray spectrum
and can therefore not determine the maximum or minimum of the energy-density spectrum \Fnu. We can only say
that the maximum is below the low-energy limit of JEM-X of 3 keV. It should be noted here
that peak values up to 5.5 keV were measured with BeppoSAX (Massaro et al. \cite{Massaro}). The spectrum
also falls off continuously to the upper energy limit of 200 keV, so a possible minimum of the \Fnu\ 
spectrum must lie at an energy greater than 200 keV.

\begin{table}
\caption{Results of fits to the Swift-BAT data with a simple power law for both states
 for a reference energy of 1 keV.}  
\label{table:7}      
\begin{tabular}{l c c c }        
\hline\hline                 
state of source & $\chi^2_{red}$ & normalization constant & $\alpha$ \\    
 &  & [keV$^{-1}$ cm$^{-2}$ s$^{-1}$] & \\
\hline                        
 quiescent & 0.681 & 0.53 $\pm$ 0.2 & 2.46 $\pm$ 0.11 \\ 
 active & 1.42 & 1.3 $\pm$ 0.4 & 2.49 $\pm$ 0.14 \\
\hline                                   
\end{tabular}
\end{table}

The data from Swift-BAT were also spectrally analysed for the quiescent and the active state.
They could be best fit with a simple power law. The results are summarised in Table~\ref{table:7}.
Within the errors, the spectral indices are identical for the quiescent and active states. This is
consistent with the results of the
data of Tables~\ref{table:4} and \ref{table:5} for the joint power law where the spectral
index also does not change with intensity. However, the spectral index obtained from the Swift-BAT
data is with \about2.5 somewhat larger than the one obtained from the JEM-X and IBIS/ISGRI data
(\about2.3) but consistent with our result within 3$\sigma$.

The intensity of the Swift-BAT data between the quiescent and active states changes by a factor of
2.45 $\pm$ 1.2. This is a bit higher than the values derived from the results of
Tables~\ref{table:4} and \ref{table:5} but still compatible within the error.

\subsection{X-Ray luminosity}

By integrating the broken power-law spectrum over the energies 3--200 keV and multiplying with the
surface of a sphere one obtains the isotropic X-ray luminosity in this energy range.
Using the parameters of Tables~\ref{table:4} and \ref{table:5} one finds
luminosities of 1.75$\cdot 10^{45}$ erg/s and 3.73$\cdot 10^{45}$ erg/s
for the quiet and active states, respectively. This is much lower than the Eddington limit,
which is in the range of (2.4 - 9.6)$\cdot$10$^{46}$ erg/s. But it should be noted that the emission
is actually beamed and that the luminosity is thus lower than given above.

\subsection{Hardness ratios}

\begin{figure}
   \centering
   \includegraphics[width=9cm]{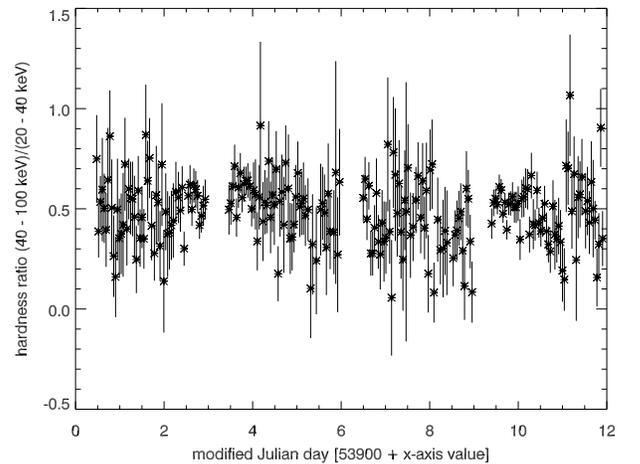}
   \caption{Hardness ratio H$/$L as a function of time for the IBIS/ISGRI data.}
   \label{ratio0}
 \end{figure}

\begin{figure}
   \centering
   \includegraphics[width=9cm]{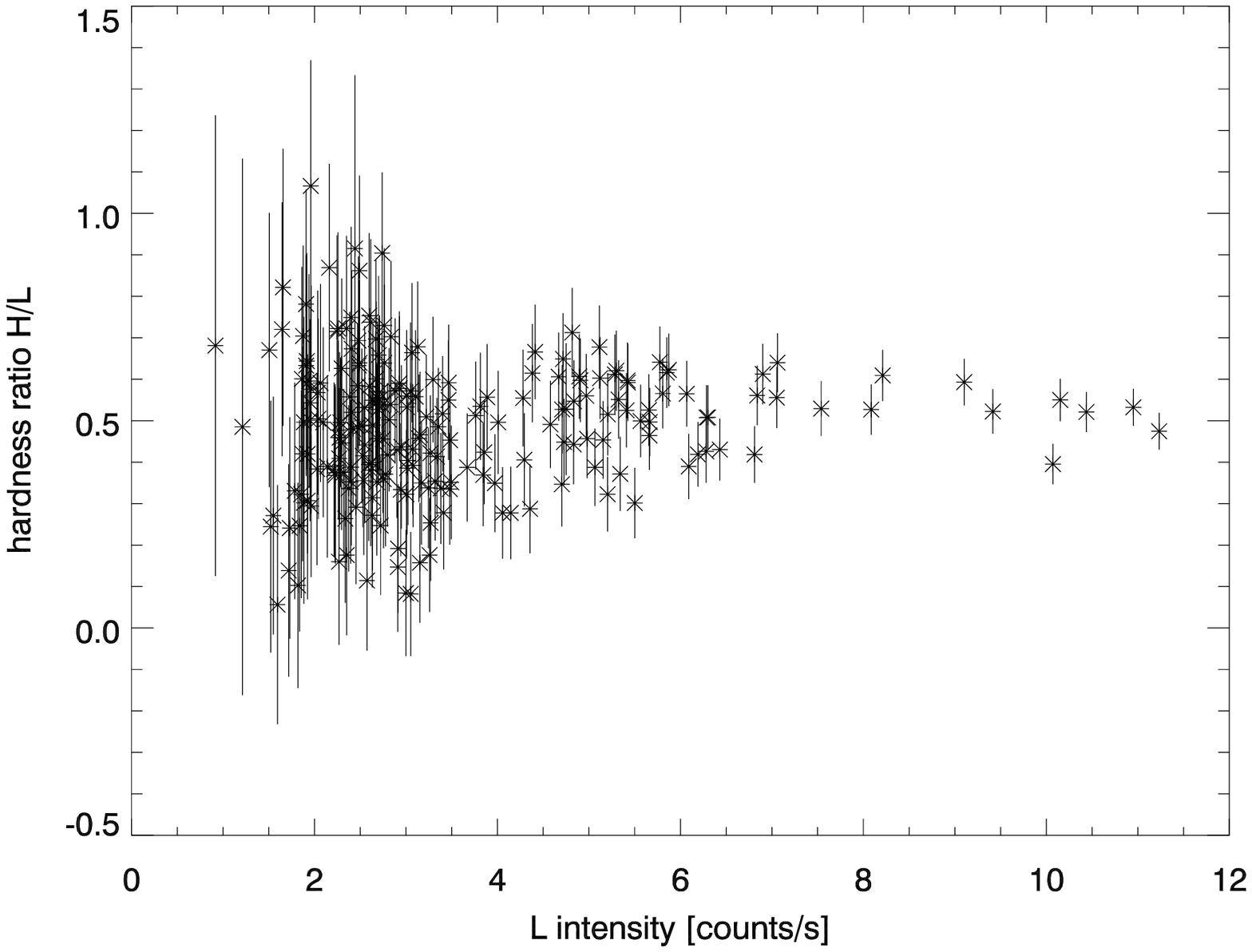}
   \caption{Hardness ratio H$/$L as a function of the L-band (20 - 40~keV) intensity for the IBIS/ISGRI data.}
   \label{ratio1}
 \end{figure}

\begin{figure}
   \centering
   \includegraphics[width=9cm]{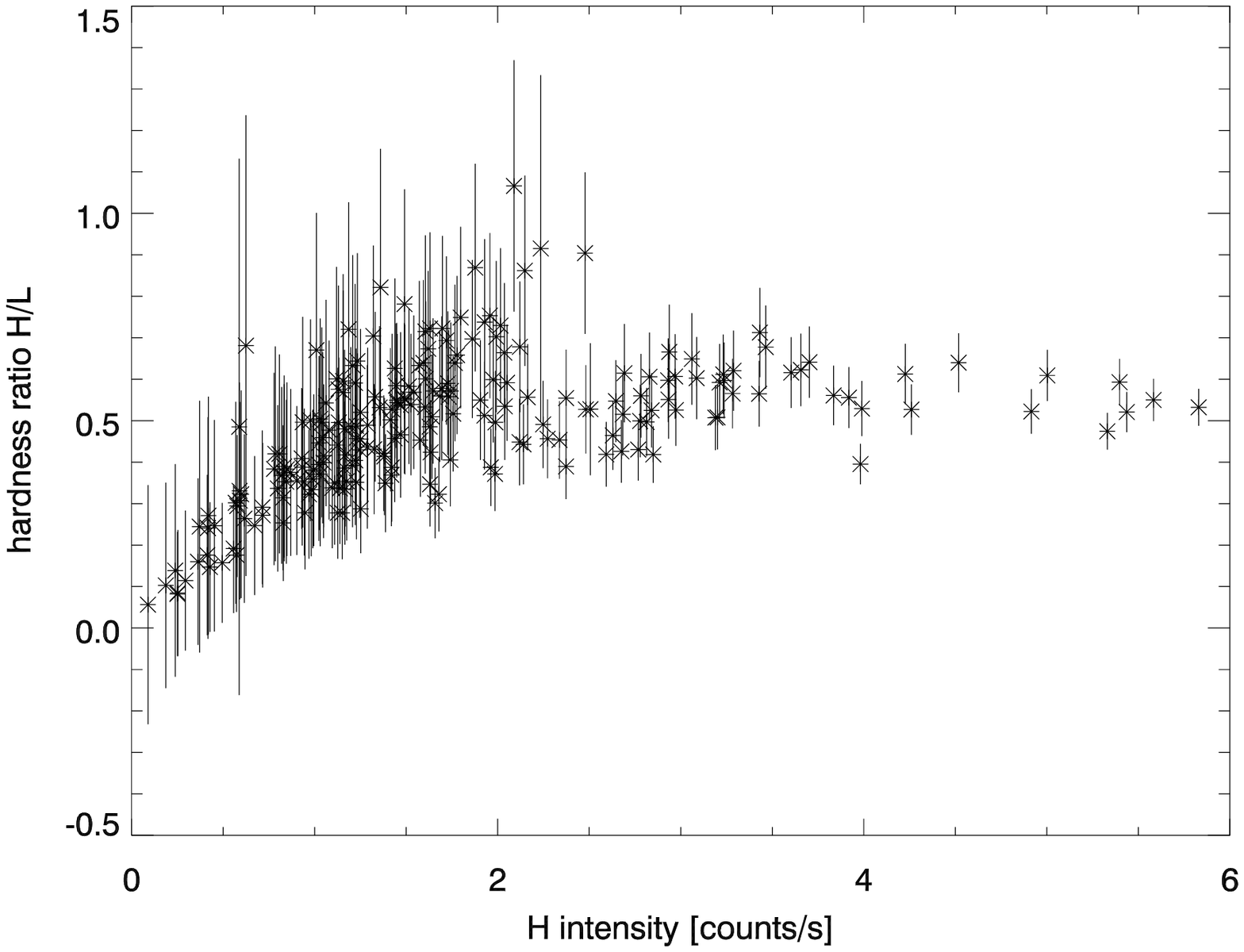}
   \caption{Hardness ratio H$/$L as a function of the H-band (40 - 100~keV) intensity for the IBIS/ISGRI data.}
   \label{ratio2}
 \end{figure}

We calculated the hardness ratio (HR) of the two IBIS/ISGRI energy bands 20 - 40 keV (L-band) and 40 - 100 keV (H-band) 
HR = H-band rate $/$ L-band rate. It is plotted in Figure~\ref{ratio0} as a
function of the time and in Figures~\ref{ratio1} and \ref{ratio2}
as functions of the L-band and the H-band intensities.
The hardness ratio in Figure~\ref{ratio0} is fairly constant in time and does not follow the
lightcurves of Figures~\ref{ISGRIlow-LC} and~\ref{ISGRIhigh-LC}, so we do not observe an evolution
in the hardness ratio with intensity.
When inspecting Figures ~\ref{ratio1} -~\ref{ratio2}, one notes a remarkable difference:
whereas no correlation is obvious when one plots the ratios as a function of the L-band count rate
(Figure ~\ref{ratio1}), a correlation for intensities up to \about 2 counts/s seems to be present
when the plot is done as a function of the H-band count rate
(Figure~\ref{ratio2}). But this might be because there are no
events with $<$1~counts/s in Figure~\ref{ratio1}.

\begin{figure}
   \centering
   \includegraphics[width=9cm]{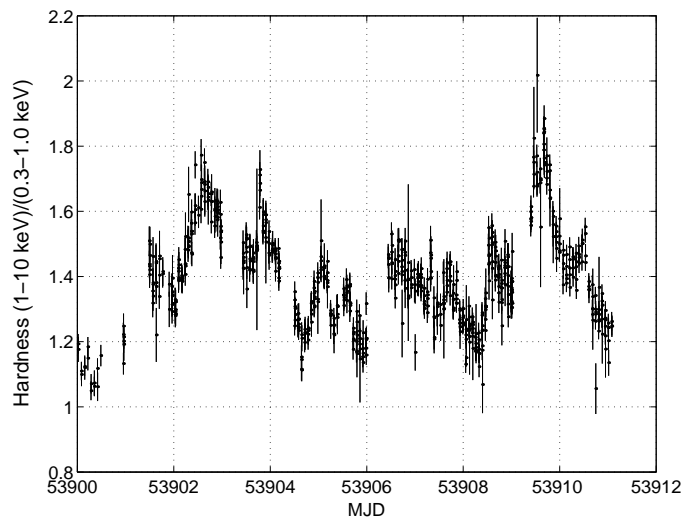}
   \caption{Hardness ratio of the Swift-XRT data as a function of the time.}
   \label{ratio5}
\end{figure}

\begin{figure}
   \centering
   \includegraphics[width=9cm]{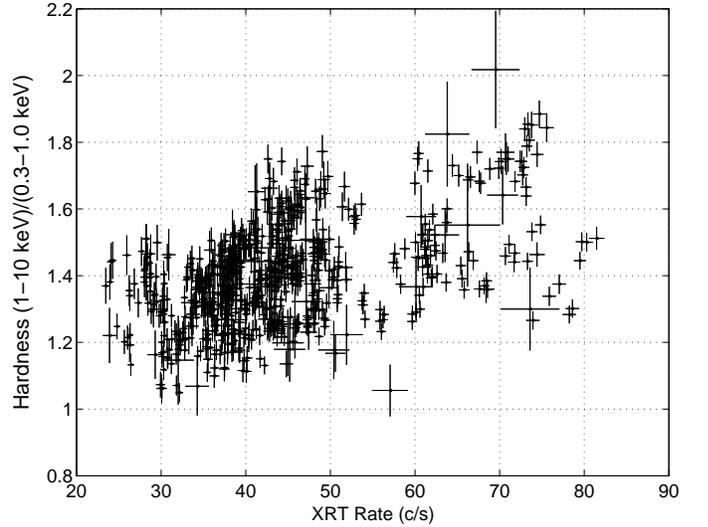}
   \caption{Hardness ratio of the Swift-XRT data as a function of the XRT intensity.}
   \label{ratio-intensity}
\end{figure}

In Figure~\ref{ratio5} the hardness ratio is shown for two energy bands [(1-10 keV)/(0.3-1 keV)] of the
Swift-XRT. A comparison with the lightcurve of Figure~\ref{XRT-LC} shows that the hardness
ratio follows the lightcurve closely. This means that we have observed an intensity-hardness-ratio
correlation; i. e., the brighter the source the harder the spectrum. But this is obviously not
valid at all times. The last peak in Figure~\ref{XRT-LC} (at MJD $\approx$ 53910.5) is only marginally visible
in Figure~\ref{ratio5}. One would expect a hardness ratio of $\sim$2 (similar to the hardness ratio
observed at MJD $\approx$ 53909.5), but actually it is only about 1.5. This means that
the emission process must have changed significantly on a time scale of $<$1 day. We obviously
observed a clear hard-to-soft evolution.

The overall hardness-ratio correlation
is also visible when one plots the Swift-XRT hardness ratio as a function
of the Swift-XRT count rate as in Figure \ref{ratio-intensity}. Although the positive correlation
with the rate is not striking it is clearly recognisable. The large scatter around this correlation
is probably due to a short-time scale and overlapping variability. 

\subsection{Multiwavelength spectrum \Fnu}

\begin{figure*}
 \centering
\includegraphics[width=18cm]{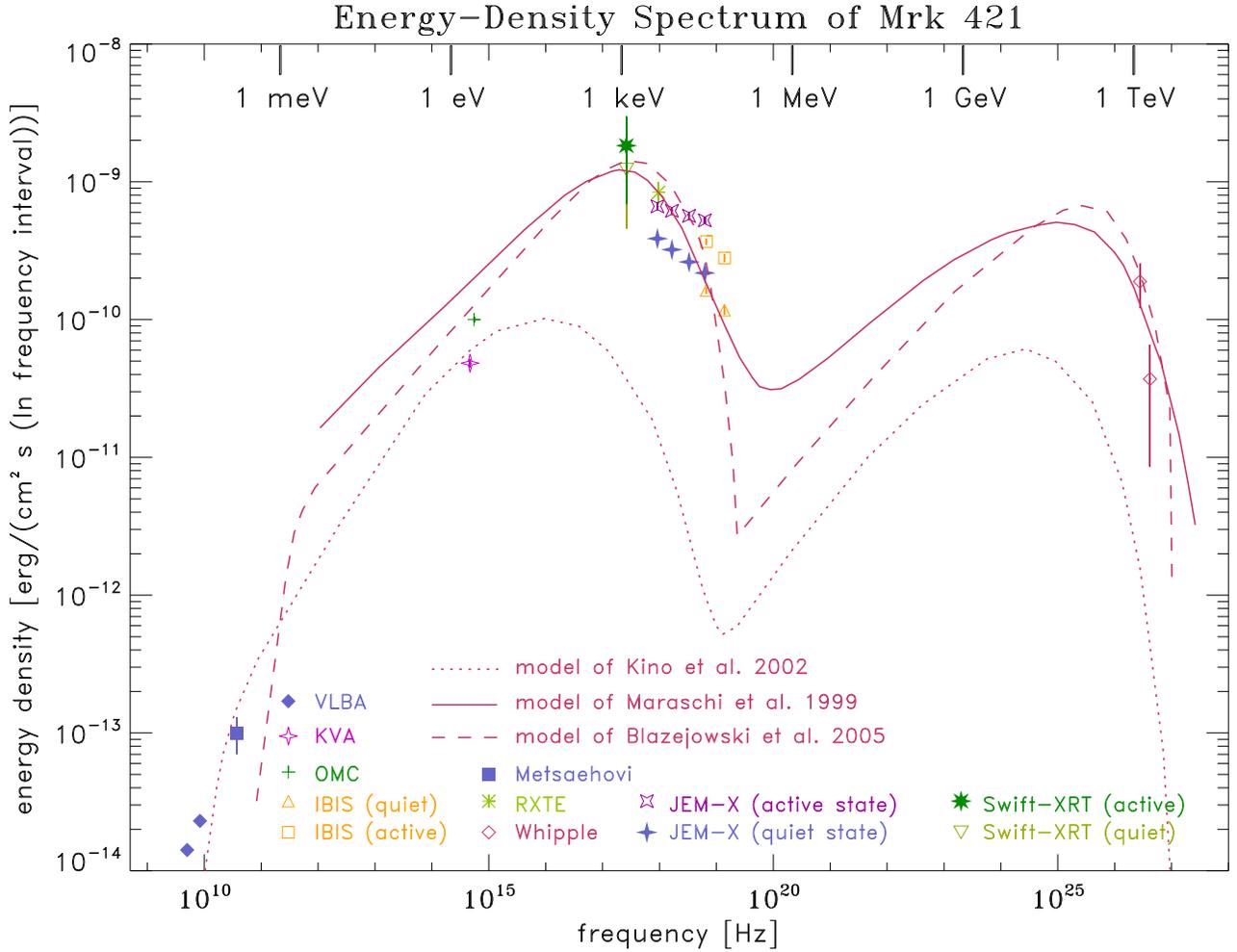}
 \caption{The measured energy-density spectrum of Mrk 421 is compared with the three theoretical models
  of\ \cite{Mar} (1999), Kino et al. (\cite{Kino}), and Blazejowski et al. (\cite{Blaz}). Please note that the
data and the three models represent the source in different states and therefore reflect its spectral
variability.}
 \label{nuFnu}
\end{figure*}

The data from all observations (with the exception of the JEM-X and ISGRI data) were averaged over the
observation time span, corrected for extinction at optical wavelengths using the extinction value E$_{B-V}$ = 0.03
of Burstein and Heiles (\cite{Burstein}) and the formalism of Seaton (\cite{Seaton}), converted to
the same unit [erg/(cm$^2$ s)], and then plotted in an energy-density spectrum (\Fnu\ spectrum).
This multiwavelength spectrum is shown in Figure \ref{nuFnu}. Unfortunately the number of data points is
sparse, especially at energies above \about 500 keV where only two data points (measured at different times)
at TeV energies exist.
It should also be noted that the data used to create the average high state and low state SED data 
points shown in Figure \ref{nuFnu} were extracted from the same time windows for 
all instruments; however, the data are not strictly simultaneous.  Due to 
this non-simultaneous data, it is possible that small time-scale flaring 
could cause a systematic offset of the SED from one energy band to another. This could explain the
discrepancy between the Swift-XRT data and the JEM-X and IBIS data.

The data are compared with the theoretical models of\ \cite{Mar} (1999), Kino et al. (\cite{Kino})
and Blazejowski et al. (\cite{Blaz}). The models of\ \cite{Mar} (1999) and Blazejowski et al.
(\cite{Blaz}) were adjusted to the flux measured by IBIS in the quiescent state by applying a factor of 3.2.
Both models predict the actual flux measured by Whipple at TeV energies quite well, but seem to fail
to fit the optical data and the radio flux. On the other hand, the model of Kino et al. (\cite{Kino}) fits the radio data quite well,
but has problems fitting the X-ray and TeV data, together with the radio data. It is remarkable that the JEM-X spectrum is flatter
than predicted by the models, even though their shape differs significantly. The three models also do not predict the
break at an energy of ~43 keV. 

The best agreement between data and models is found for the model of Blazejowski et al. (\cite{Blaz}).
However, it should be noted here that the models were fitted to data measured at different times, so it is not surprising
that they do not fit our data well. In contrast, they show that not only the intensity, but also the spectral shape of Mrk 421
is extremely variable.

\subsection{Model fits}

\begin{figure*}
 \centering
 \includegraphics[width=18cm]{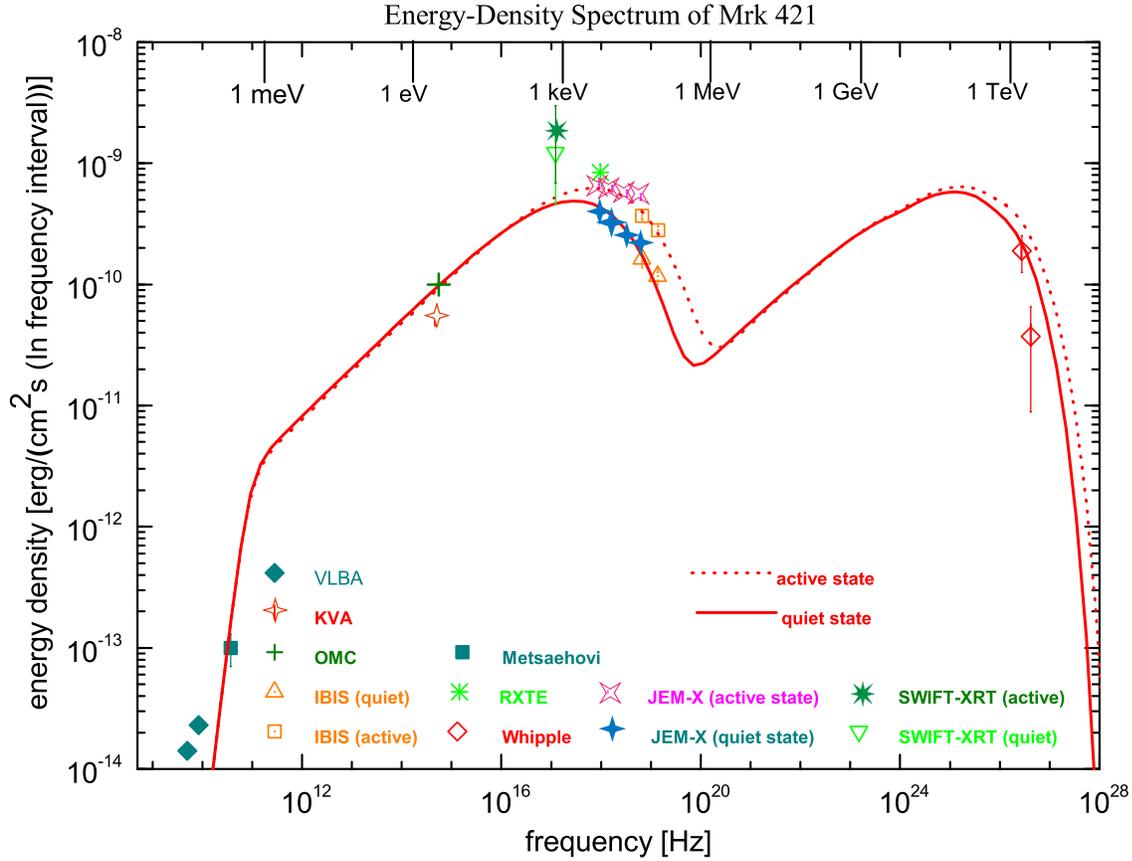}
 \caption{The result of the model fits to the measured energy-density spectrum of Mrk 421.}
 \label{modelspectrum}
\end{figure*}

We also tried to fit a theoretical emission model to the data of the energy-density spectrum of
Figure \ref{nuFnu}. The model curves fitting the data shown in Figure \ref{modelspectrum} best were
obtained using a one-zone synchrotron self-Compton model (SSC) including the
full Klein-Nishina cross section for inverse Compton scattering (\cite{Jones},
\cite{Blumenthal}). The model assumes a spherical
blob with constant injection of non-thermal electrons and a constant
escape rate for electrons and photons. The model details are described
in \cite{Rueger} (2007).

In this calculation the electron distribution is assumed to be constanst
(stationary) and is represented by a smoothed broken power law combined with an
exponential cut-off as given in \cite{Tavecchio01}:

\begin{equation}
N(\gamma)=K \gamma^{-s_1}\left(1+\frac{\gamma}{\gamma_b}\right)^{s_1-s_2} \exp \left(- \frac{\gamma}{\gamma_{max}}\right)
\label{elec}
\end{equation}

\noindent with the differential electron number density $N$, Lorentz factor
$\gamma$, break energy $E_b=\gamma_b m c^2$, spectral indices $s_1$
and $s_2$, normalization factor $K$, and cut-off energy $\gamma_{max}$.
This phenomenological Ansatz for the spectrum is justified by
self-consistent simulations (see \cite{Rueger} 2007). 

With this electron distribution, the radiative transfer equation was
solved numerically in the comoving frame and the observed spectrum
obtained by applying the beaming effect using the Doppler-factor
$\delta$. Specifically, the transformation for the intensity is

\begin{equation}
I_{\nu, beamed}=\delta^3 \,I_{\nu, comoving frame}.
\end{equation}

In addition, one can show that, for dominating synchrotron losses, the second
spectral index can be rewritten as $s_2=s_1+1$, so one less parameter
has to be determined. A first estimation of the parameter set is
obtained by using the equations given in \cite{Kataoka2}. Small
variations lead to the best-fit parameter set presented in Table
\ref{table:para}.

\begin{table}
\caption{Best-fit parameters for high-state and low-state intensity when
a one-zone synchrotron self-Compton model is fit to the spectrum.}
\label{table:para}
\centering
\begin{tabular}{ccc}
\hline\hline
  parameter      &    high-state     & low-state \\
\hline   
 $\gamma_b$      & $4 \cdot 10^5  $  & $3 \cdot 10^5$ \\
 $\gamma_{max}$  & $1.2 \cdot 10^6$  & $7 \cdot 10^5$ \\
 $s_1$           & 2.2               & 2.2 \\
 $s_2$           & 3.2               & 3.2 \\
 $B$ [G]         & 0.1               & 0.1 \\
 $K$ [cm$^{-3}$] & $1.6 \cdot 10^4 $ & $1.6 \cdot 10^4 $ \\
 $\delta$        & 15                & 15 \\
\hline
\end{tabular}
\end{table}

From this table one can see that both parameter sets differ only in the break energy
$\gamma_b$ and the cut-off energy $\gamma_{max}$. The latter energy
represents the efficiency of the acceleration mechanism. Since $\gamma_b$
results from the emerging balance between cooling and acceleration of
the electrons, a more efficient acceleration mechanism leads to
both a higher $\gamma_{max}$ and a higher $\gamma_b$. Therefore we
conclude that the variability of the emission can be due to varying
efficiency of the acceleration.

It should be noted, however, that a change in the spectrum of the injected
electrons without changing the environmental parameters like the magnetic fields and/or the
Doppler factor is difficult to understand. But the acceleration of particles can be influenced
without changing the environmental parameters by taking turbulence into account which can have
a dramatic influence on the particle acceleration; on the one hand, enhanced turbulence can change
the spectral indices (\cite{VainioSchlick} and \cite{VainioSpa}), and on the other, the decay length of turbulence can
change the maximum energy (\cite{VainioSchlick} and \cite{VainioSpa}). However these effects are not covered in our model,
just like the geometry of the shock and the shock thickness.

The fit presented here is not unique (see discussion in \cite{Tavecchio02}) and the observed change
in spectral energy distribution can also be modelled by slightly varying the magnetic field strength
and Doppler factor. The changes in these parameters would lead to change in electron spectrum, and
such detailed modelling is beyond the scope of this paper. The Doppler factor and magnetic-field strength
modelled here are similar to those derived for other TeV-emitting blazars. However, very long baseline
interferometric observations show that such high Doppler factors would require very extreme assumptions
for the viewing angle (0.015 degrees) and the Lorenz factor (25) (\cite{Piner}). Therefore
adopting models where the jet has a velocity structure (\cite{Georganopoulos}, \cite{Ghisellini})
might be more feasible, but such modelling is beyond the scope of the paper.

\section{Discussion and conclusions}

The blazar Mrk 421 underwent an active phase in May/June 2006. INTEGRAL was therefore reoriented
to observe this source for 12 days with the instruments OMC, JEM-X, and IBIS.
Simultaneously the source was also observed by the Mets\"ahovi and VLBA radiotelescopes,
by the KVA telescope in the optical waveband, by RXTE and SWIFT at X-rays, and by Whipple at TeV
energies. In X-rays, several strong flares were observed not seen at lower energies.
Unfortunately the data at TeV energies are too sparse to allow such a conclusion for the high
energies. In this context it is especially striking that the flares are not visible in the optical
lightcurves, since it is thought that the X-rays and optical photons are produced by the same
electron population via synchrotron emission. One would therefore expect similar lightcurves;
but when fitting a theoretical one-zone synchrotron self-Compton model to the data, it is found
that flares seen in X-rays may not necessarily be visible at lower frequencies.

However, other possibilities also exist for explaining the observed behaviour, rather than varying the efficiency
of the acceleration. For example, it could be that the optical photons originate in larger emission regions and
that the flares are thus occurring  on longer time scales and are therefore smeared out. It could 
also be that the intensity fluctuations seen in X-ray are damped in the optical band. This view could be supported
by the observed intensity correlation between OMC and IBIS in Figure~\ref{Figure4}. Whereas the IBIS rate varies by a
factor of  $\sim$5, the variation in the optical flux is only a few percent, so the high variations seen in X-rays are strongly
damped in the optical range, hence barely visible. This could be achieved, for example, if the optical photons
were obscured by some intervening clouds of matter that are transparent for the X-rays but not for the optical photons.

The spectral analysis of the JEM-X and IBIS/ISGRI data gave a surprising result:
the spectral index $\alpha$ of the power law for the two states does not change much (from 2.12 to 2.3)
at energies below ~40 keV, although the intensity changes by factors up to 5. In the energy range
2--10 keV, \cite{Takahashi} found a clockwise evolution of the spectral index as a function of the
intensity. But in their case the range of the change of the spectral index is nearly the same as
reported by us (from 2.3 to 2.52), although the intensity only changes by a factor of ~2.
This shows that the observed slight hardening occurs predominantly at low intensities.
This ''saturation effect'' is clearly seen in Figure~\ref{ratio2}.

Another remarkable result is that the hardness ratio of the Swift-XRT data is strongly varying
with intensity (see Figure~\ref{ratio5}), whereas the hardness ratio of the IBIS/ISGRI data is more or less
constant (see Figure~\ref{ratio0}). This again shows that there is a conspicuous difference in the emission processes of the
low- and high-energy X-rays. At energies $<$10 keV, the spectrum becomes harder with increasing intensity.
This is not the case for X-rays with energies $>$20 keV. In the standard model in which the X-rays are
produced by synchrotron radiation from the same electron population, this different behaviour is difficult
to explain.

The picture becomes even more complex when one compares the lightcurves of Figure~\ref{XRT-LC}
to Figure~\ref{ISGRIhigh-LC}. The two flares at the end of the observation interval (around MJD 53910)
show a noticeable spectral difference: the first peak is much harder than the second one!
If one assumes that each flare is produced by a relativistically-moving blob of electrons, the physical
characteristics of each electron population and their environment (i. e. magnetic fields etc.)
must be different, so the picture of jet emission is very complex and the extraction of the relevant
information from the data requires a detailed and profound modelling of the emission processes.

\begin{acknowledgements}

We thank Tal Alexander for providing his program for the calculation of the ZDCF and for making
his paper about this program available to us prior to publication. We also express our
gratitude to Albert Domingo Garau who assisted us in analysing the OMC data correctly.
A. Falcone and D. Morris acknowledge support from NASA contract NAS5-00136. 
Part of this work was supported by the "Ministerium f\"ur Bildung und Forschung" via the DLR under
the grant 50.OG.9503.0. The Mets\"ahovi team acknowledges the support from the Academy of Finland.
The Very Long Baseline Array (VLBA) is a facility of the The National Radio-Astronomy Observatory (NRAO),
which is operated by Associated Universities, Inc., under cooperative agreement with the National Science Foundation. 
We also thank an anonymous referee for her/his useful and valuable comments.

\end{acknowledgements}

{}

\end{document}